\documentclass[oldversion]{aa} 
\usepackage{graphicx}
\usepackage{aalongtable}
\usepackage{txfonts}
\usepackage{rotating}
\usepackage{lscape}
\newcommand{\gsim}{\;\lower.6ex\hbox{$\sim$}\kern-7.75pt\raise.65ex\hbox{$>$}\;}
\newcommand{\lsim}{\;\lower.6ex\hbox{$\sim$}\kern-7.75pt\raise.65ex\hbox{$<$}\;}

\begin{document}
\title{Na-O Anticorrelation and HB. VII. The chemical composition of first and
second-generation stars in 15 globular clusters from GIRAFFE 
spectra\thanks{Based on observations  collected at ESO telescopes under
programmes 072.D-507 and 073.D-0211}
 }

\author{
E. Carretta\inst{1},
A. Bragaglia\inst{1},
R.G. Gratton\inst{2},
S. Lucatello\inst{2,3},
G. Catanzaro\inst{4},
F. Leone\inst{5},
M. Bellazzini\inst{1},
R. Claudi\inst{2},
V. D'Orazi\inst{2},
Y. Momany\inst{2,6},
S. Ortolani\inst{7},
E. Pancino\inst{1},
G. Piotto\inst{7},
A. Recio-Blanco\inst{8},
\and
E. Sabbi\inst{9}
}

\authorrunning{E. Carretta et al.}
\titlerunning{Na-O anticorrelation in 15 globular clusters}

\offprints{E. Carretta, eugenio.carretta@oabo.inaf.it}

\institute{
INAF-Osservatorio Astronomico di Bologna, Via Ranzani 1, I-40127
 Bologna, Italy
\and
INAF-Osservatorio Astronomico di Padova, Vicolo dell'Osservatorio 5, I-35122
 Padova, Italy
\and
Excellence Cluster Universe, Technische Universit\"at M\"unchen, 
 Boltzmannstr. 2, D-85748, Garching, Germany 
\and
INAF-Osservatorio Astrofisico di Catania, Via S.Sofia 78, I-95123 Catania, Italy
\and
Dipartimento di Fisica e Astronomia, Universit\`a di Catania, Via S.Sofia 78, I-95123 
Catania, Italy
\and
European Southern Observatory, Alonso de Cordova 3107, Vitacura, Santiago, Chile
\and
Dipartimento di Astronomia, Universit\`a di Padova, Vicolo dell'Osservatorio 2,
I-35122 Padova, Italy
\and
Laboratoire Cassiop\'ee UMR 6202, Universit\`e de Nice Sophia-Antipolis, CNRS,
Observatoire de la Cote d'Azur, BP 4229, 06304 Nice Cedex 4, France
\and
Space Telescope Science Institute, 3700 San Martin Drive,
Baltimore, MD 21218, USA
  }

\date{Received 18 March 2009 / Accepted 31 May 2009}

  \abstract{We present abundances of Fe, Na, and O for 1409 red giant stars in 15
  galactic globular clusters (GCs), derived from the homogeneous analysis of
  high-resolution FLAMES/GIRAFFE spectra. Combining the present data with 
  results from our FLAMES/UVES spectra and from previous studies within the project,
  we obtained a total sample of 1958 stars in 19 clusters,
  the largest and most homogeneous database of this kind to date. 
  The programme clusters cover a range  in metallicity from [Fe/H]$=-2.4$ dex to [Fe/H]$=-0.4$
  dex, with a wide variety of global parameters (morphology of the
  horizontal branch, mass, concentration, etc.). For all clusters we find the
  Na-O anticorrelation, the classical signature of the operation of proton-capture
  reactions in H-burning at high temperature in a previous generation of more
  massive stars that are now extinct. Using quantitative criteria (from the morphology
  and extension of the Na-O anticorrelation), we can define three different
  components of the stellar population in GCs. We separate a primordial
  component (P) of first-generation stars, and two components of second-generation stars, that we name   intermediate (I) and extreme (E) populations
  from their different chemical composition. The P component is present in all
  clusters, and its fraction is almost constant at about one third. The I
  component represents the bulk of the cluster population. On the other hand, E
  component is not present in all clusters, and it is more conspicuous in some (but
  not in all) of the most massive clusters. We discuss the fractions and spatial
  distributions of these components in our sample and in two additional clusters
  (M~3=NGC~5272 and M~13=NGC6205) with large sets of stars analysed in the literature. We also find
  that the slope of the anti-correlation (defined by the minimum O and maximum
  Na abundances) changes from cluster-to-cluster, a change that is represented well
  by a bilinear relation on cluster metallicity and luminosity. This second
  dependence suggests a correlation between average mass of
  polluters and cluster mass.
  }

\maketitle

\section{Introduction}

At the turn of XXI century, the notion of GCs as true
examples of simple stellar populations had to face a serious challenge. 
Astrophysicists realised that the long standing idea of complete chemical homogeneity 
among  stars within a cluster only applies to nuclei forged in
core-collapse or thermonuclear supernovae (iron-group elements and the heaviest
of the $\alpha-$elements).  On the other hand, lighter elements like C, N, O,
Na, Mg, Al, and F (for which abundance measurements in GC stars were obtained only recently, 
e.g. Smith et  al. 2005) show large star-to-star abundances variations. 
This pattern is clearly different from
what observed among  field stars in the same evolutionary 
stages, where only C and N (and Li) abundances are
observed to change,  while the abundances for the remaining light elements only
reflect a typical pattern of supernova nucleosynthesis: field stars only
populate a well-defined region at (constant at a 
given [Fe/H]\footnote{We adopt the usual spectroscopic notation, $i.e.$ 
[X]= log(X)$_{\rm star} -$ log(X)$_\odot$ for any abundance quantity X, and 
log $\epsilon$(X) = log (N$_{\rm X}$/N$_{\rm H}$) + 12.0 for absolute number
density abundances.}) high O, low Na abundances.

Some years ago, the most popular explanation for the cluster 
stars peculiar compositions involved some degree of internal mixing due to the stars
evolving along the red giant branch (RGB: see the review by Kraft 1994). However,
it is  currently well-established that, even if a certain degree of evolutionary
mixing is  present both in field (Gratton et al. 2000) and cluster (Smith and
Martell 2003)  stars, its impact is confined to Li, C, and N. 
The explanation for the observed star-to-star variations in the abundances 
of heavier nuclei, usually found to be
anti-correlated (O and Na, Mg and Al), and even for the observed variations of CH
and CN band strength in cluster turn-off stars (e.g.  Cannon et al. 1998, Briley
et al. 2004), had to be looked for elsewhere.

The key observation was the detection by Gratton et al. (2001) 
among unevolved stars in NGC~6752 and NGC~6397 of Na, O
variations, anti-correlated with each other. 
This observation, confirmed afterward in other clusters (M~71=NGC~6838, Ramirez
and Cohen 2002; 47 Tuc=NGC~104, Carretta et al. 2004a), definitively ruled out the
possibility that the abundance variations are generated by processes occurring
inside observed stars,  because of the rather low central temperatures and thin
convective envelopes of stars at the turn-off of GCs. 

The scenario currently accepted invokes an external origin for the abundance
variations, very likely the pollution from matter enriched with elements cycled
through proton capture H-burning reactions at high temperature (Denisenkov \&
Denisenkova 1989, Langer et al. 1993) of the intra-cluster gas from which the
stars, that we presently observe, did form out (see Gratton, Sneden and Carretta 2004
for a recent review).

This scenario requires that more than one stellar generation formed
within each GC. It is very likely that this is the normal
succession of events leading to the formation of these aggregates, since
abundance variations are observed in each GC studied to date. However, 
the class of stars playing the role of major, early polluters cannot 
be established yet (e.g. fast rotating massive stars, Decressin et al. 
2007; or intermediate-mass AGB stars, D'Antona and Ventura 2007 and 
references therein). What is clear is
that the old definition of ``abundance anomalies" can be dropped, and 
the more modern issue of the chemical composition and nature of second
generation stars in GCs should be addressed.

This is the seventh paper in a series aimed at studying the mechanisms of
formation and early evolution of stellar generations in GCs and, by
investigating the relations between their properties and the global cluster
parameters, the scenarios for the formation of the GCs themselves. The project
is named {\it Na-O anticorrelation and horizontal branch (HB)}, its main
emphasis being the possible link between the compositions of stars along the RGB and
the HB morphology in GCs.

Such a connection has been suspected for a long time, with the He abundances
as a {\it trait d'union}: He enhancement in cluster stars was invoked both by
theoretical predictions of yields from rotating massive stars and intermediate
mass asymptotic giant branch (AGB) stars and by photometric 
observations showing in some cases HBs with
extremely long blue tails and multiple sequences (see the review by Cassisi et
al. 2008 and Piotto 2009 for references and recent updates). The bottom line is
that He-enhanced stars are likely to populate the blue extreme of the HBs and to
also explain the extreme O-depletion observed in the surface abundances of RGB
stars.

To better quantify the relation between the chemical composition of first/second
generation stars and HB morphology, we started homogeneous analysis of the
FLAMES spectra for more than 2000 stars in 19 GCs with different metallicity, HB
morphology, and global parameters (mass, age, density, etc.).

The plan and the general strategy of our project has already been explained in
the first paper of the series, so we briefly summarise it here for the convenience 
of the reader.
Carretta et al. (2006a, hereafter Paper I) was
dedicated to NGC~2808, the classical template for a bimodal distribution of HB 
stars. While explaining the tuning of the analysis procedures and tools for
dealing with hundreds of stars in a large sample of GCs, previous papers were devoted to 
the study of
particular objects. In Carretta et al (2007a, Paper II) we analysed
NGC~6752, a cluster with a long blue HB and a relatively modest extension of the
Na-O anticorrelation. Three papers (Gratton et al. 2006, 2007, and Carretta et
al. 2007b, Paper III, V, and VI, respectively), focused on the two peculiar
bulge clusters NGC~6441 and NGC~6388. Paper IV (Carretta et al. 2007c) dealt with
the analysis of the Na-O anticorrelation in NGC~6218 and the first detection of
a He-poor/He-rich stellar population among giant stars in GCs.

The collection and analysis of all the observational material is now
complete and this unprecedented database of
abundance ratios can be used to gain new insight into the formation processes leading to
the GCs that we currently observe after nearly a Hubble time.

The aims of the present paper are three-fold: first,  we present results from
GIRAFFE spectra for the remaining  15 GCs in our sample,
homogeneously deriving Fe, O, and Na abundances for about 1500 stars. Second, we
combine the results from previous papers, to have the  full set of observed
Na-O anticorrelation in all 19 GCs from FLAMES/GIRAFFE. Third, data obtained
from FLAMES/UVES spectra\footnote{Except for NGC~6441 and  NGC~6388, already
published in Papers III and VI, the analysis of the UVES spectra is described in
a companion paper, Carretta et al. (2009), hereinafter Paper VIII.} will be
merged with the GIRAFFE dataset  to improve statistics and discuss on solid grounds 
the chemical composition of different stellar generations in GCs
and to highlight their basic properties.

Inferences on cluster evolution and correlations with global cluster parameters
derived from the present data will be thoroughly discussed in two forthcoming
papers (Carretta et al. in preparation, Gratton et al. in preparation). The last
two columns in Table~\ref{t:qualigc} summarise for clarity the references to the
papers where the analysis of all GIRAFFE and UVES data is  presented for each
cluster in our project.

The paper is organised as follows: an outline of the target selection criteria
and observations is given in Section 2, the derivation of atmospheric 
parameters and the analysis are described in Section 3,  error estimates
are briefly discussed in Sect. 4. In Section 5 we show the Na-O anticorrelation in
all clusters and identify  different components among the stellar
populations in GCs, based on their chemical composition. The Na content of first
and second-generation stars is discussed in Section 6. A dilution model for the
Na-O anticorrelation is sketched in Section 7. Finally, a summary is presented in
Section 8. In the Appendix a more detailed discussion of the procedure followed to
estimate star-to-star and cluster errors is given.

\section{Target selection and observations}

Our foremost aim is to systematically and fully explore any possible
connection between the chemical signature of different stellar populations in
GCs and the distribution of stars in the colour-magnitude diagram (CMD) during the HB
phase. We tried therefore to target GCs with the widest variety of HB
morphologies. 

We then selected clusters with a (stubby) red HB (NGC~104, NGC~6171, NGC~6838), 
with an HB populated from red to blue colours (NGC~3201, NGC~4590, NGC~5904,
NGC~6121, NGC~7078) and with a predominantly blue HB (NGC~288, NGC~1904, NGC~6218,
NGC~6254, NGC~6397, NGC~6752, NGC~6809, NGC~7099); some objects show a {\it
very} extended blue HB  (NGC~1904, NGC~6218, NGC~6254, NGC~6752, NGC~7078).
Finally, three clusters with bimodal distributions in the HB (NGC~6388, NGC~6441,
and NGC~2808, all also showing very extended blue HBs) were included among our
targets.

In Table~\ref{t:qualigc} some useful information are listed (Galactocentric radius,
foreground reddening, apparent visual distance modulus, HB type, and metallicity
[Fe/H]), taken from the updated online version of the catalogue by Harris (1996).
 In our sample we have clusters with metal abundances from
[Fe/H]$=-2.4$ to about [Fe/H]$=-0.4$, spanning almost the whole metallicity
range of the  galactic GCs.

\begin{figure}
\centering
\includegraphics[bb=25 160 330 710, clip, scale=0.69]{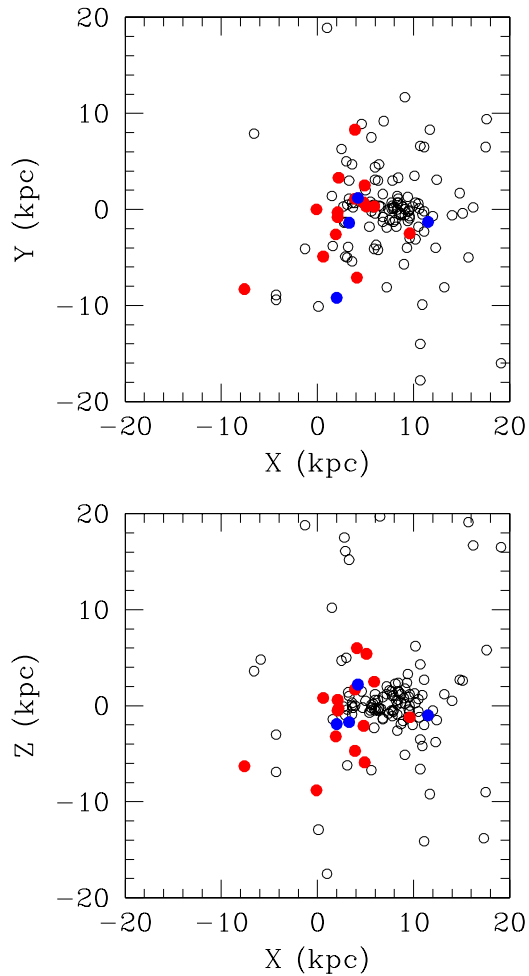}
\caption{Location of out target clusters in a Sun-centred coordinate system,
where X points toward the Galactic centre,
Y in the direction of Galactic rotation and Z toward the North Galactic Pole. 
Distance components are in kiloparsec. Filled red circles are GCs analysed in
the present work and filled blue circles the GCs already published in this
project, superimposed on all clusters in the Harris's (1996) database.}
\label{f:xyz}
\end{figure}

Figure~\ref{f:xyz} shows the location of our target GCs in a
Sun-centred coordinate system\footnote{X points toward the Galactic centre, Y
in the direction of Galactic rotation and Z toward the North Galactic Pole. 
Distance components are in kiloparsec.}, superimposed on all clusters in the 
Harris's (1996) database. Due to observational constraints, the clusters in our
sample are, whenever  possible, those lying nearer to the Sun's location. However,
apart from this obvious limitation, there is nothing peculiar in the spatial
distribution of our sample (corresponding to about 13\% of the $\sim $150
known GCs in the Galaxy) with respect to the location of the other clusters.

The clusters can be grouped  for age and kinematical properties according to the
classical division introduced by Lee, Zinn and co-workers, whose latest and more
complete compilation is from Mackey and van den Bergh (2005). We observed four
so-called bulge/disc clusters (NGC~104, NGC~6388, NGC~6441, NGC~6838), 12
objects in the old halo group (NGC~288, NGC~1904, NGC~2808, NGC~5904, NGC~6121,
NGC~6171, NGC~6218, NGC~6254, NGC~6397, NGC~6752, NGC~6809, NGC~7099) and three
in the young halo subgroup (NGC~3201, NGC~4590, NGC~7078).  Finally, the range
in mass covers more than one order of magnitude, from M~71 (NGC~6838, absolute magnitude
$M_V = -5.60$ (Harris 1996) up to NGC~6441 ($M_V=-9.64)$. It is noteworthy
that five out of the nine most massive GCs in our Galaxy are in our
sample. Summarizing, on the basis of
 Table~\ref{t:qualigc} and Figure~\ref{f:xyz} we can be
reasonably certain that our sample is representative of the global GC
population, with no particular bias and/or selection effects.

The spectroscopic data were collected in service mode using the ESO 
high-resolution multifibre spectrograph  FLAMES/GIRAFFE (Pasquini et al. 2002)
mounted on the VLT UT2. Observations were done with two GIRAFFE setups, the
high-resolution gratings HR11 (centred on 5728~\AA) and HR13 (centred on
6273~\AA), which were respectively chosen to measure the Na doublets at
 5682-5688~\AA\ and 6154-6160~\AA\ and the [O {\sc i}] forbidden lines at 6300,  6363~\AA, as well as
several lines of Fe-peak and $\alpha-$elements. The spectral resolutions are
R=24,200 (for HR11) and R=22,500  (for HR13), at the centre of the spectra. Total
exposure times  obtained for each cluster are listed in Table~\ref{t:qualigc}.
The average seeing during the observations was less than 1.1 arcsec.

\begin{figure*}
\centering
\includegraphics[scale=0.50]{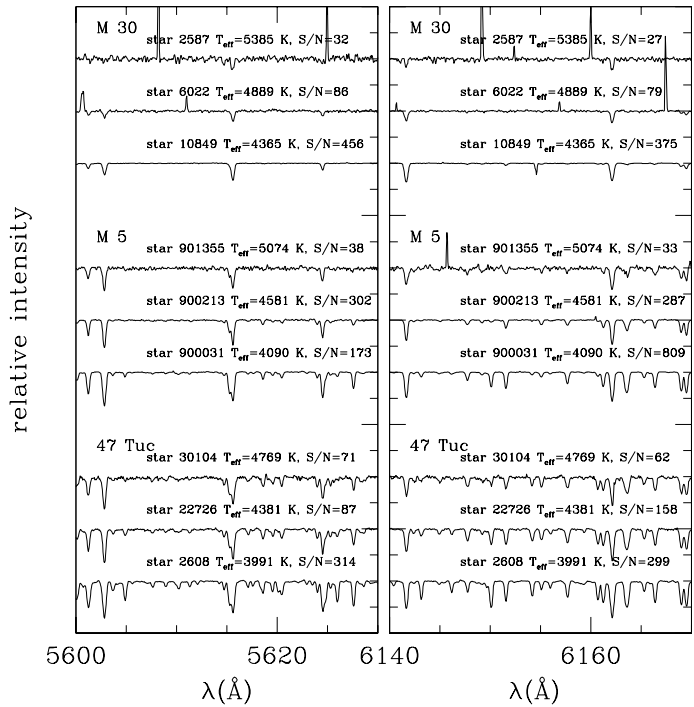}
\caption{Examples of observed spectra obtained with FLAMES/GIRAFFE and the HR11
(left panel) and HR13 (right panel) gratings. Displayed are portions of spectra
of three stars in a metal-rich (47~Tuc=NGC~104), a metal-intermediate
(M~5=NGC~5904), and a
metal-poor cluster (M~30=NGC~7099). The stars are at the middle and at the extremes of
the range in temperature (magnitude) sampled in each clusters. Spectra are
normalised to the continuum and shifted by arbitrary quantities for display
purposes. The effective temperatures and the S/N are indicated for
each star.}
\label{f:spettri}
\end{figure*}

In Figure~\ref{f:spettri} we show a few examples of the spectra acquired with
FLAMES/GIRAFFE and the HR11 and HR13 gratings in one metal-rich (47~Tuc=NGC~104), one
metal-intermediate (M~5=NGC~5904), and one metal-poor cluster (M~30=NGC~7099). For each cluster 
we displayed stars in the middle and at the ends of the sampled range in
temperature (magnitude), with typical $S/N$\ per pixel. As is evident also
from this figure, the $S/N$\  is not a simple linear function of the
magnitude, due to the different throughput of the fibres, and to slightly
different degrees of precision in the centreing of the targets in each fibre.

\begin{table*}
\caption{Main characteristics of the target clusters and references.}
\small
\begin{tabular}{llrllclcrrll}
\hline
GC       &      &$R_{GC}$&E(B-V)& (m-M)v & HBR & [Fe/H]& range M$_V$         & Texp.& Texp.&Giraffe & UVES\\
         &      &        &      &        &     &       &                     & (sec)&(sec) &        &     \\        
         &      &        &      &        &     &       &                     & HR11 & HR13 &        &     \\
\hline
NGC  104 &47~Tuc&  7.4   &0.04& 13.37& -0.99& -0.76& $-$1.1 $\div$   +1.2&  3200&  1600 &Paper VII &Paper VIII \\
NGC  288 &      & 12.0   &0.03& 14.83&  0.98& -1.24& $-$1.7 $\div$   +1.6& 10800&  5400 &Paper VII &Paper VIII \\
NGC 1904 &M~79  & 18.8   &0.01& 15.59&  0.89& -1.57& $-$2.3 $\div$   +1.5& 11700& 11700 &Paper VII &Paper VIII \\
NGC 2808 &      & 11.1   &0.22& 15.59& -0.49& -1.15& $-$1.7 $\div$ $-$0.1&  8850& 11700 &Paper I   &Paper VIII \\
NGC 3201 &      &  8.9   &0.23& 14.21&  0.08& -1.58& $-$0.9 $\div$   +2.5&  3600&  3600 &Paper VII &Paper VIII \\
NGC 4590 &M~68  & 10.1   &0.05& 15.19&  0.17& -2.06& $-$0.8 $\div$   +2.4&  7200& 10200 &Paper VII &Paper VIII \\
NGC 5904 &M~5   &  6.2   &0.03& 14.46&  0.31& -1.27& $-$1.8 $\div$   +1.6&  4100&  4100 &Paper VII &Paper VIII \\
NGC 6121 &M~4   &  5.9   &0.36& 12.83& -0.06& -1.20& $-$1.2 $\div$   +1.2&   950&   950 &Paper VII &Paper VIII \\
NGC 6171 &M~107 &  3.3   &0.33& 15.06& -0.73& -1.04&   +0.6 $\div$   +2.4&  8100& 10800 &Paper VII &Paper VIII \\
NGC 6218 &M~12  &  4.5   &0.19& 14.02&  0.97& -1.48& $-$2.0 $\div$   +1.6&  2700&  2700 &Paper IV  &Paper VIII \\
NGC 6254 &M~10  &  4.6   &0.28& 14.08&  0.98& -1.52& $-$1.2 $\div$   +1.8&  2800&  2800 &Paper VII &Paper VIII \\
NGC 6388 &      &  3.2   &0.37& 16.14& -0.70& -0.60& $-$0.8 $\div$   +1.6& 31400& 39100 &Paper VII &Paper VI   \\
NGC 6397 &      &  6.0   &0.18& 12.36&  0.98& -1.95& $-$1.3 $\div$   +2.4&   900&   900 &Paper VII &Paper VIII \\
NGC 6441 &      &  3.9   &0.47& 16.79& -0.70& -0.53& $-$0.6 $\div$   +0.3& 10600& 10600 &Paper V   &Paper III  \\
NGC 6752 &      &  5.2   &0.04& 13.13&  1.00& -1.56& $-$1.3 $\div$   +1.4&  1750&  1750 &Paper II  &Paper VIII \\
NGC 6809 &M~55  &  3.9   &0.08& 13.87&  0.87& -1.81& $-$2.5 $\div$   +1.5&  4100&  2200 &Paper VII &Paper VIII \\
NGC 6838 &M~71  &  6.7   &0.25& 13.79& -1.00& -0.73& $-$0.2 $\div$   +1.2&  2700&  2700 &Paper VII &Paper VIII \\
NGC 7078 &M~15  & 10.4   &0.10& 15.37&  0.67& -2.26& $-$2.6 $\div$   +1.6&  8100&  8100 &Paper VII &Paper VIII \\
NGC 7099 &M~30  &  7.1   &0.03& 14.62&  0.89& -2.12& $-$2.0 $\div$   +2.6&  5400&  5400 &Paper VII &Paper VIII \\
\hline
\end{tabular}
\label{t:qualigc}
\\
Galactocentric distance, coordinates, foreground reddening, apparent visual
distance modulus, horizontal branch ratio HBR=(B-R)/(B+V+R), and metallicity from
the catalogue by Harris (1996) and web updates.

\end{table*}

As done for the previous GCs (Papers I to VI), our targets were selected among
isolated stars near the RGB ridge line\footnote{All stars were chosen to be
free from any companion closer than 2 arcsec and brighter than $V+2$ mag, where
$V$ is the target magnitude.}. To reduce concerns related to model atmospheres
and ensure the sampling of sufficiently populated regions of the CMD, 
stars close to the  RGB tip were generally avoided.

The number of actual cluster members observed 
in each cluster (as well as the typical
$S/N$  of the spectra) depends on several factors: 
\begin{itemize}
\item[a)] the size of the cluster red giant population, which in turn depends
mainly on the cluster mass and somewhat on the cluster distance (more massive/distant 
clusters allowed/forced us to observe brighter stars, less massive/more nearby 
clusters required/allowed us to shift down to fainter stars to
gather enough targets to fully exploit the maximum number of
dedicated fibres);
\item[b)] the area covered by the cluster on the sky, which depends on its
distance and concentration; for objects with smaller angular sizes we encountered more
severe problems in positioning the FLAMES fibres (using the dedicated tool FPOSS).
Hence, in the case of NGC~288 we observed stars at only 0.07 core radii
from the cluster centre; in the opposite situation, for the highly concentrated
and distant NGC~6388, the first sampled distance from the centre is about 17.5
core radii (but for every cluster all fibres were placed within its tidal radius);
\item[c)] field stars contamination: this problem is exacerbated in particular
for disc/bulge clusters such as NGC~6171, NGC~6388, NGC~6441, and NGC~6838. For
these objects a somewhat limited number of member stars was observed. Moreover,
we were forced to reject a number of potential target candidates in the most
metal-poor clusters (e.g. NGC~7099, NGC~7078, NGC~4590) where the very small
number of (usually weak) lines hampered the assessment of the membership and the
abundance analysis.
\end{itemize}

The approximate range in absolute $V$ magnitude for stars observed in each
cluster is given in Table\ref{t:qualigc}. For several GCs this range extends
down to luminosities fainter than the level of the RGB-bump.

We used the available optical photometry  calibrated to the standard 
Johnson-Cousins system  (Landolt 1992) for our target selection. 
The published photometric data
are from Bellazzini et  al. (2001) for NGC~288;  Momany et al. (2003) for
NGC~4590, NGC~7078, and NGC~7099\footnote{Data for NGC~104, NGC~6121, and NGC~6171 were 
not
published, but were nevertheless reduced exactly like the others in 
Momany et al. (2003)};  Momany et al. (2004) for NGC~1904,
and NGC~7099. Details on the other unpublished  photometric catalogues are beyond the 
purpose of the present discussion, so we provide some brief information
for reference.
 Clusters NGC~5904, NGC~6254, NGC~6397, and NGC~6809 were observed
with  the Wide Field Imager (WFI, FoV $33\arcmin \times 32\arcmin$), mounted on
the 2.2m ESO/MPI  telescope in La Silla, Chile. For NGC~5904, B, V images were
obtained with short (5 sec) and long  (200-400 sec) exposures on UT 2000 July 7.
The sky conditions were not optimal, with clouds and bad seeing, so the WFI 
photometry was only used to complement (in area) the B, V photometry by
Sandquist et al. (1996), and was calibrated by  comparison. For NGC~6254 the
photometry is obtained from a couple of V and I  images with exposure time 4 min
and a couple of V and I images with 10 seconds. Instrumental magnitudes were 
obtained with Dophot (Schechter et al. 1993) and transformed into the standard
 Johnson/Kron-Cousins system using 84 secondary  standard stars from the Stetson
(2000) set that were in common with the cluster catalogue. Photometry for
NGC~6397 and NGC~6809 consists in short (3-4 seconds  and 5-8 seconds,
respectively, for NGC~6397 and NGC~6809) and long (70-90 seconds and 90-180,
respectively) V and B images  (proposal 69.D-0582, P.I. Ortolani). For these two
clusters, data were  reduced using Daophot II (Stetson 1994) in {\sc
IRAF}\footnote{IRAF is  distributed by the National Optical Astronomical
Observatory, which are operated by the Association of Universities for Research
in Astronomy, under contract with the National Science Foundation }, and
calibrated to the standard system. For NGC~3201 and NGC~6838, we adopted
unpublished  photometry kindly provided by C. Corsi and L. Pulone (private
communication). For each cluster we used the Guide Star catalogue (GSC-2) to 
search for astrometric standards in the entire WFI image field of view. Several
hundred astrometric GSC-2 reference stars  were found in each chip, allowing
us an accurate absolute position of the detected stars ($\sim$ 0.2 arcsec r.m.s.
in  both R.A. and Dec.). Finally, photometric and astrometric data for NGC 6388
are described in Paper VI.

A list of all the GIRAFFE target spectra retained in our final sample,
together with  coordinates, magnitudes, and radial velocities (RVs), is given in
Table~\ref{t:coo} (the full table is only available in electronic form at CDS). 
Together with the stars in the previously published clusters, and those
with UVES spectra from Paper VIII, the number of objects with abundances derived
from intermediate or high-resolution spectra is 1958. The project database
 increases by {\it an order of magnitude} the total number of RGB
stars with abundance analysis in galactic GCs (the literature
samples up to now consisted of a total of about 200 stars scattered among
several clusters).  Moreover, our abundance analysis is as homogeneous as 
currently possible, for the procedures for measuring equivalent widths
($EW$s), derivation of atmospheric parameters, list of atomic parameters, and set
of model atmospheres.

Field stars (established on the basis of their radial velocities) were disregarded and
excluded from further analysis if the measured RV differed by more than
3$\sigma$ from the cluster average. In some cases, cross check of membership
with available proper motions was possible (M4: Cudworth \& Rees, 1990; M~5: 
Cudworth, 1979; NGC~6171: Cudworth et al. 1992; M~71: Cudworth 1985;  M15:
Cudworth 1976) and used to further clean out the member list. Contamination from
stars on the AGB is only a minor source of concern
for our analysis, because a priori the occurrence of AGB stars is expected to 
be at most about 10\% of that of RGB stars. Moreover, the RGB and AGB are
usually well separated at the luminosity of the observed stars. A posteriori,
the very small scatter in derived iron abundances in each cluster  ensures that
we are using reliable atmospheric parameters, including the adopted stellar mass
(appropriate for RGB).

We used the 1-D, wavelength-calibrated spectra as reduced by the dedicated
Giraffe pipeline (BLDRS  v0.5.3, written at the Geneva Observatory, see {\em
http://girbldrs.sourceforge.net}). Radial velocities were  measured using the
{\sc IRAF} package {\sc FXCORR} with appropriate templates and are shown in
Table~\ref{t:coo}. 

Since we also aimed to target up to 14 stars per cluster with the dedicated 
UVES fibres (see Paper VIII), the GIRAFFE fibre positioning
between the HR11 and HR13 pointings had to be changed. Because of this, 
not all the stars were observed with
both gratings. Among a total of 1409 {\it bona fide} cluster members  observed
with GIRAFFE, 765 have spectra with both gratings, 320 only have HR11 
observations, and 324 only HR13 observations.  While we could recover Na
abundances even for stars only observed with HR13 (at least for metal-rich
clusters), since the weaker Na doublet at 6154-6160~\AA\ falls into the spectral
range covered by this setup, we could expect to measure oxygen for only a
maximum of 1089 stars.

\begin{table*}
\centering
\caption{List and relevant information for the 1409 target stars   The complete
table is available electronically only at CDS.}
\begin{tabular}{rrrrcccrrrl}
\hline
\hline
GC      &ID    &RA           &Dec           &$B$    &$V$    &$I$   &$K$	    &RV(HR11) &RV(HR13)&Notes       \\
\hline
NGC~104 & 1389 & 0 24  7.423 & -71 56 56.67 & 14.855&13.847 &0.000 & 11.099 &         & -16.63 &  HR13      \\
NGC~104 & 2608 & 0 25  0.617 & -71 55 58.66 & 13.654&12.250 &0.000 &  8.617 & -26.73  & -26.97 &  HR11,HR13 \\
NGC~104 & 2871 & 0 24 40.034 & -71 55 45.03 & 14.950&13.983 &0.000 & 11.321 & -20.73  &        &  HR11      \\
NGC~104 & 4373 & 0 23 18.186 & -72 11 51.64 & 15.292&14.345 &0.000 & 11.978 & -10.94  & -11.51 &  HR11,HR13 \\
NGC~104 & 5172 & 0 23  9.787 & -72 11 18.38 & 14.861&13.823 &0.000 & 11.292 & -18.01  & -18.48 &  HR11,HR13 \\
\hline
\end{tabular}
\label{t:coo}
\end{table*}

\section{Atmospheric parameters and analysis}

\subsection{Atmospheric parameters}

Temperatures and gravities were derived using the same procedure we described in 
the previous papers of the series (see Papers I to VI);  along with the
derived microturbulent velocities and iron abundances, they are listed in
Table~\ref{t:atmpar} (completely available only in electronic form at CDS) for
all the 1409 stars having GIRAFFE spectra in the 15 clusters analysed in this
work.

\begin{table*}
\centering
\caption[]{Adopted atmospheric parameters and derived iron abundances. The
complete Table is available electronically only at CDS.
}
\begin{tabular}{rrccccrcccrccc}
\hline
GC &Star   &  $T_{\rm eff}$ & $\log$ $g$ & [A/H]  &$v_t$	     & nr & [Fe/H]{\sc i} & $rms$ & nr & [Fe/H{\sc ii} & $rms$ \\
   &    &     (K)	&  (dex)     & (dex)  &(km s$^{-1}$) &    & (dex)	  &	  &    & (dex)         &       \\
\hline
NGC~104  & 1389 &4568 &2.09 &-0.78 &1.66 & 21 &-0.775 & 0.131 & 2  & -0.724 & 0.208 \\  
NGC~104  & 2608 &3991 &0.99 &-0.77 &1.64 & 44 &-0.770 & 0.159 & 3  & -0.748 & 0.084 \\  
NGC~104  & 2871 &4609 &2.17 &-0.74 &1.10 & 19 &-0.737 & 0.106 &    &        &       \\  
NGC~104  & 4373 &4709 &2.38 &-0.80 &1.42 & 44 &-0.800 & 0.198 & 4  & -0.732 & 0.170 \\  
NGC~104  & 5172 &4560 &2.08 &-0.71 &1.35 & 40 &-0.711 & 0.137 & 3  & -0.753 & 0.029 \\  
 \hline
\end{tabular}
\label{t:atmpar}
\end{table*}

Effective temperatures ($T_{\rm eff}$) were obtained in two steps. We derived
first estimates of $T_{\rm eff}$ and bolometric corrections (B.C.) for our stars
from $V-K$ colours, where $V$ is from our photometry and $K$ was taken from the
Point Source Catalogue of 2MASS (Skrutskie et al. 2006) and transformed to the
TCS photometric system, as used in Alonso et al. (1999). We employed the
relations by Alonso et al. (1999, with the erratum of 2001). For all
clusters the  distance moduli, values of foreground reddening, input
metallicities as listed in Table~\ref{t:qualigc} (Harris 1996) were adopted, and the
relations  $E(V-K) = 2.75 E(B-V)$, $A_V = 3.1 E(B-V)$, and $A_K = 0.353 E(B-V)$
(Cardelli et al. 1989). We checked that the use of more recent relations between
monochromatic absorption and reddening, like those of Fitzpatrick (1999),
including dependence of reddening corrections on stellar colours, has
negligible impact in our analysis, with differences in the temperatures $<10$~K.

In the second step, as in Paper II and the subsequent  papers of this project,
the final adopted  $T_{\rm eff}$ were derived from a relation between $T_{\rm
eff}$ (from $V-K$ and the Alonso et al. calibration) and $V$ or $K$ magnitude.
To derive this relation, we used ``well-behaved" stars in each cluster (i.e.
stars with magnitudes in both visual and infrared filters and lying on the RGB).
This procedure was adopted to decrease the  scatter in abundances due
to uncertainties in temperatures, since magnitudes are much more reliably
measured than colours. The assumptions behind this approach are discussed in
Paper II to which we refer the reader for details.

Surface gravities log $g$ were obtained from the apparent magnitudes, the  above
effective temperatures and distance moduli, and the  bolometric corrections from
Alonso et al. (1999), assuming  masses of  0.85 M$_\odot$\footnote{We note that
the derived values of surface gravity are not very sensitive to the exact value
of the adopted mass}  and  $M_{\rm bol,\odot} = 4.75$ as the bolometric
magnitude  for the Sun.
As usual, we derived values of the microturbulent velocities $v_t$'s by
eliminating trends in the relation between abundances from Fe neutral lines  and
expected line strength (see Magain 1984).

Final metallicities were then obtained by interpolating, in the Kurucz (1993) 
grid of model atmospheres (with the option for overshooting on), the model with 
the proper atmospheric parameters whose abundance matches that derived from  Fe
{\sc i} lines.

\subsection{Equivalent widths and iron abundances}

Adopted line lists, atomic parameters, and reference solar abundances (from Gratton et al. 2003)
are strictly homogeneous for all stars analysed in the present programme.
 Equivalent widths ($EW$s) were measured as described in
detail in  Bragaglia et al. (2001) with the same automatic procedure we used in
the previous analysis of GIRAFFE spectra (Papers I, II, VI, V) for the 
definition of the local continuum around each line. This is a crucial step at
the limited resolution of our spectra, especially for the coolest targets. 

\begin{figure}
\centering
\includegraphics[scale=0.45]{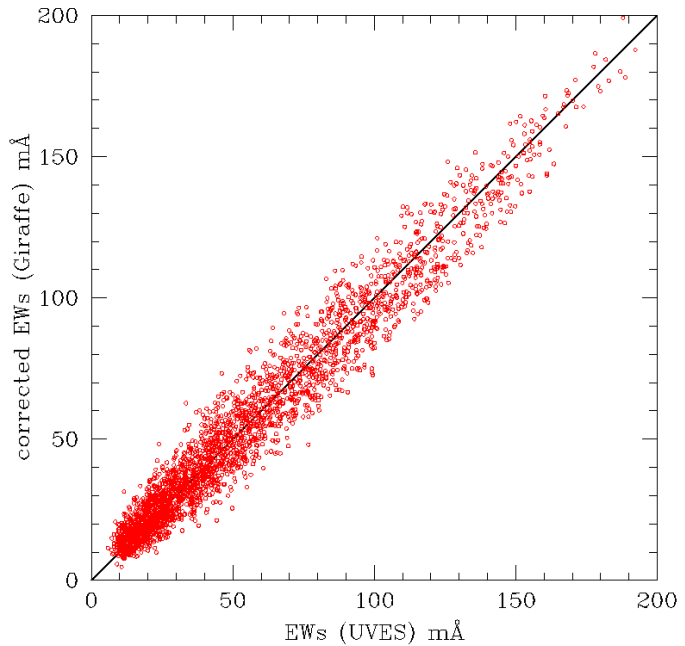}
\caption{Comparison between the $EW$s measured on high-resolution FLAMES/UVES
spectra and those measured for the same stars on the GIRAFFE spectra, after they
were corrected to the system of the UVES $EW$s (see text).}
\label{f:corretteEW}
\end{figure}

As in the previous papers, we corrected the $EW$s measured in the
intermediate-resolution GIRAFFE spectra to the system defined by the 
high-resolution UVES spectra, using the stars observed with both instruments in each
cluster (see Paper VIII). This correction was deemed necessary since the contribution of unrecognised blends can cause an overestimate of the $EW$s measured on intermediate resolution spectra.  On the other hand, veiling from very weak lines, again not recognizable on lower resolution spectra, might lower the true continuum, resulting into an underestimate of measured $EW$s.

In 13 out of 15 clusters we had a number of stars observed with both instruments,
 from a minimum of five up to 13 stars, with an average of about 10 stars 
 per cluster. However, in
NGC~1904 and NGC~6838, no stars in common between the UVES and GIRAFFE samples were
available. In the first case we used five UVES stars with 
{\em a relative difference} in effective temperature  within 10 K from five
GIRAFFE stars for our comparison: since the cluster does not show any large  intrinsic scatter in
element ratios (obviously, with the exceptions of Na, O, Mg, and Al lines), this is a
reasonable approach. In the case of NGC~6838, the target stars of UVES
observations are much cooler than those observed with GIRAFFE and a similar
comparison is impossible. To correct the $EW$s in this cluster we then applied
the average relation derived from the other  13 GCs.
Figure~\ref{f:corretteEW} shows the comparison between the $EW$s measured on UVES
spectra and the corrected $EW$s from GIRAFFE spectra. After this correction the
average difference (in the sense UVES minus GIRAFFE) is $+0.1\pm 0.2$~m\AA\
($rms=8.1$~m\AA) from 2811 lines.

\begin{table*}
\centering
\caption{Average iron abundances from UVES (from Paper VIII) and GIRAFFE 
spectra.}
\begin{tabular}{lcccclrrr}
\hline
GC       &[Fe/H]  &[Fe/H]I$\pm$stat.err.& syst.&$rms$ &N. stars &[Fe/H]II & $rms$ & N.stars \\
         &UVES    &GIRAFFE            &error  &       &          &         &       &         \\
         & (dex)  &  (dex)            & (dex) &  (dex)&          &  (dex)  &	 (dex) &	  \\
\hline
NGC  104 &$-$0.768 &$-$0.743$\pm$0.003 & $\pm$0.026 & 0.032 &147 & $-$0.769& 0.075 & 110 \\
NGC  288 &$-$1.305 &$-$1.219$\pm$0.004 & $\pm$0.070 & 0.042 &110 & $-$1.231& 0.092 &  72 \\
NGC 1904 &$-$1.579 &$-$1.544$\pm$0.005 & $\pm$0.069 & 0.036 & 58 & $-$1.483& 0.061 &  50 \\
NGC 3201 &$-$1.512 &$-$1.495$\pm$0.004 & $\pm$0.073 & 0.049 &149 & $-$1.403& 0.106 &  99 \\
NGC 4590 &$-$2.265 &$-$2.227$\pm$0.006 & $\pm$0.068 & 0.071 &122 & $-$2.233& 0.108 &  10 \\
NGC 5904 &$-$1.340 &$-$1.346$\pm$0.002 & $\pm$0.062 & 0.023 &136 & $-$1.348& 0.072 & 109 \\
NGC 6121 &$-$1.168 &$-$1.200$\pm$0.002 & $\pm$0.053 & 0.025 &103 & $-$1.197& 0.056 &  80 \\
NGC 6171 &$-$1.033 &$-$1.065$\pm$0.008 & $\pm$0.026 & 0.044 & 33 & $-$1.053& 0.085 &  26 \\
NGC 6254 &$-$1.575 &$-$1.556$\pm$0.004 & $\pm$0.074 & 0.053 &147 & $-$1.558& 0.091 & 102 \\
NGC 6388 &$-$0.441 &$-$0.406$\pm$0.013 & $\pm$0.028 & 0.078 & 36 & $-$0.351& 0.158 &  29 \\
NGC 6397 &$-$1.988 &$-$1.993$\pm$0.003 & $\pm$0.060 & 0.039 &144 & $-$1.985& 0.077 &  32 \\
NGC 6809 &$-$1.934 &$-$1.967$\pm$0.004 & $\pm$0.072 & 0.044 &156 & $-$1.933& 0.060 & 111 \\
NGC 6838 &$-$0.832 &$-$0.808$\pm$0.005 & $\pm$0.048 & 0.034 & 39 & $-$0.801& 0.065 &  39 \\
NGC 7078 &$-$2.320 &$-$2.341$\pm$0.007 & $\pm$0.067 & 0.061 & 84 & $-$2.352& 0.091 &  27 \\
NGC 7099 &$-$2.344 &$-$2.359$\pm$0.006 & $\pm$0.067 & 0.046 & 64 & $-$2.289& 0.085 &  14 \\
\hline
\end{tabular}
\label{t:quantefe}
\end{table*}

\begin{figure}
\centering
\includegraphics[scale=0.44]{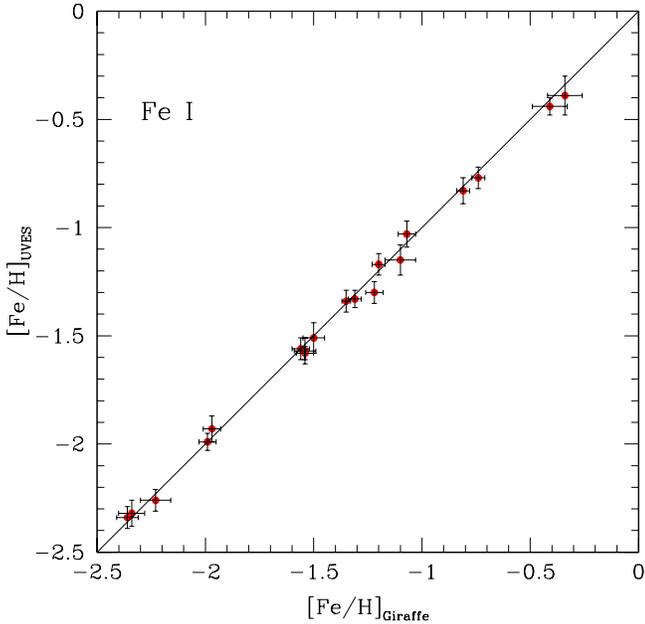}
\caption{Metal abundances obtained from GIRAFFE spectra compared with
[Fe/H]~{\sc i} ratios derived from high-resolution UVES spectra for programme GCs.
In this plot we also included the 4 clusters (NGC~2808, NGC~6752, NGC~6218, and
NGC~6441) already analysed in previous papers. Error bars are 1$\sigma$ $rms$
scatter.}
\label{f:FeI}
\end{figure}

Average abundances of iron for the 15 programme clusters derived from our 
GIRAFFE spectra, are listed in Table~\ref{t:quantefe}. As a comparison, average 
metallicities derived from the analysis of UVES spectra (Paper VIII) are
reported in the second column of this table. The agreement is very good, with the
average difference $0.007\pm0.008$ dex with $rms=0.033$ dex. We are
practically on the same  scale, as also demonstrated in Figure~\ref{f:FeI}, where
we included the four clusters previously analysed in this series. This check
is relevant, since in the following we  merge results for [O/Fe] and [Na/Fe]
ratios obtained from the samples of stars observed with both UVES
and GIRAFFE.

Metal abundances ([Fe/H]) obtained from the analysis of GIRAFFE spectra are
listed in the form [Fe/H] $\pm err1\pm err2$ dex, where the first error refers to
the statistical errors and the second one is relative to the cluster or
systematic error (see Appendix A). The $rms$ scatter and the number of stars
used in the averages are given in columns 5 and 6 of Table~\ref{t:quantefe}. 
The last 3 columns concern the abundances of iron derived from the singly
ionised  species; generally, the two averages agree very well, although the $rms$
scatter associated to the [Fe/H]{\sc ii} abundance ratio is higher. We point out
that the number of useful Fe~{\sc ii} lines in the spectral range covered by
HR11 and HR13 is very limited, at most 1 or 2. Moreover, we remind the reader
that one of the criteria in the star selection was to choose stars as far away
as possible from the tip of RGB to avoid concerns related to continuum placement
and remain in the temperature regime where model atmospheres are more reliable.
Hence, lines of Fe~{\sc ii} are not
strong for these rather warm, high-gravity stars, 
and the effect is exacerbated for clusters at very low metallicity.

The agreement we found is a good sanity check, since  the ionisation equilibrium
for Fe is quite sensitive to any possible problem in the abundance analysis,
whereas the differences we obtained are almost negligible.

\begin{figure}
\centering
\includegraphics[scale=0.44]{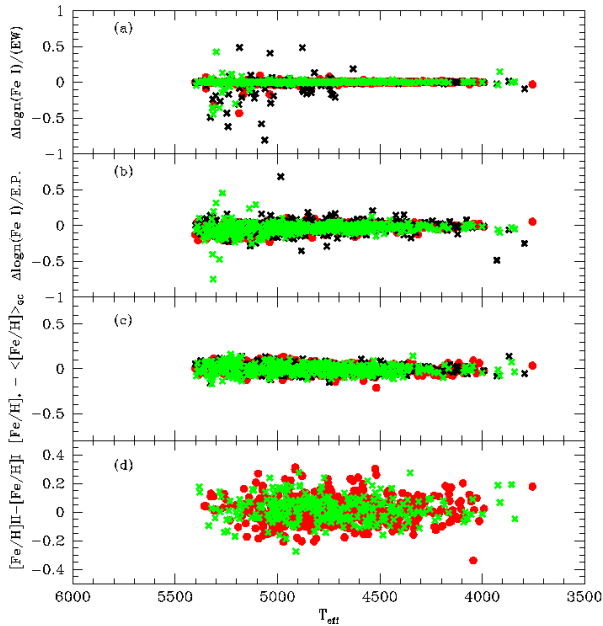}
\caption{Diagnostic diagrams for the analysis of 1409 stars with GIRAFFE 
spectra in the 15 GCs analysed here. $(a)$: slope of the relation  between
expected line strength and Fe~{\sc i} abundances used to derive the $v_t$ values
for each individual stars. $(b)$ slope of the relation between Fe~{\sc i}
abundances and excitation potential E.P. for each star. $(c)$: the average
[Fe/H] value for each cluster is subtracted from the metallicity of each star in
the cluster, and the differences are shown. $(d)$: differences in iron
abundances from Fe~{\sc i} and Fe~{\sc ii} lines. All quantities are plotted as
a function of the effective temperature. Filled red circles indicate stars with
observations in both HR11 and HR13 gratings; green and black crosses, 
are for stars observed only with HR13 and HR11, respectively.}
\label{f:analisiFe}
\end{figure}

Other diagnostic diagrams are shown in Figure~\ref{f:analisiFe}. In the upper
panel, the final slope in the relation of the expected line strength vs Fe~{\sc
i} abundances {\it for each of the 1409 individual stars} with GIRAFFE spectra
in the 15 clusters is plotted as a function of temperature, coded according to
the gratings. Apart from very few stars (mainly some warm and metal-poor stars
observed with HR11 only where there are just a few Fe~{\sc i} 
lines\footnote{In these cases we chose not to force the zeroing of the relation
Fe~{\sc i} abundances $vs$ line strength, due to the associated large 
uncertainties in the resulting fit because of the very few lines available.}), 
most slopes are near zero: the average value is $0.000 \pm0.000$ $rms=0.004$
(1293 stars), after a 2.5$\sigma-$clipping to exclude outliers. 

Panel (b) in Figure~\ref{f:analisiFe} displays the slopes of the relation between
abundances from neutral Fe~{\sc i} lines and excitation potential for each
analysed  star, as a function of the effective temperature adopted. After 126
outliers in the plot are eliminated in a 2.5$\sigma-$clipping, the average value
for this slope turns out to be $-0.023 \pm0.001$ with $rms=0.043$ (1403 stars).
In turn, this implies that on average the temperatures we derive from colours
are higher than those we would derive from the excitation equilibrium by about
80 K.

If we plot this slope as a function of the metallicity of individual stars, we
see that the difference increases with decreasing metallicity.
 A possible explanation  is that at low metallicities we
are seeing a more marked influence of departures from the LTE assumption, and/or
an atmospheric structure not reproduced well by one-dimensional model
atmospheres, as suggested by Asplund et al. (1999). Both effects are likely to be more
relevant in low metallicity stars, where the atmosphere is more transparent.

In the panel (c) we show the difference for each star between the individual
[Fe/H]~{\sc i} value and the average value for the cluster, in order to plot in
the same plane all stars in GCs of different metallicities. These differences run 
flat across a temperature range of about 1600 K, the average difference being
$-0.001 \pm 0.001$, $rms=0.041$ dex (1480 stars, again after a
2.5$\sigma-$clipping). Finally, the lower panel in Figure~\ref{f:analisiFe}
illustrates the good agreement between iron abundances from Fe~{\sc i} and
Fe~{\sc ii} species over the whole range in temperature and cluster
metallicities (after culling out 34 outliers, the average value is  $+0.016 \pm
0.002$, $rms=0.072$ dex, from 868 stars.)

\section{Errors in the atmospheric parameters and cosmic spread in Iron}

The error estimate in abundance analysis is often a poorly explained issue. In
most cases there is a certain degree of confusion between internal errors,
systematic errors and sensitivities of abundances to changes in the atmospheric
parameters. In some cases, only the last quantities are given in the papers,
with no actual estimate of errors on the derived abundances.

The procedure for error estimates perfected in previous papers of this series is
purposely tailored to deal with the approach we used to obtain the atmospheric
parameters required for the analysis. In particular, we emphasise the two steps
followed: first, we derived first-guess temperatures  from $V-K$ colours, less
sensitive to the metal abundances than other colour indices. Second, we then
derived the final adopted $T_{\rm eff}$'s from a relation between temperature on
the Alonso et al. scale and magnitude, under the assumptions (verified in each
case) that the stars  involved all belong to the RGB and that there is no intrinsic
spread in abundances in the cluster. This second step (when using the infrared
$K$ magnitudes from 2MASS) greatly alleviates problems in clusters with high and
likely differential values of the reddening, as demonstrated by the small
$rms$ scatters we obtain in the iron distributions even in GCs  (e.g. NGC~6388,
NGC~6254) well known for being affected by this phenomenon.  Moreover, using
this relation results in a sharp decrease in the star-to-star errors, since
magnitudes are more easily measured than colours, in particular for our rather
bright programme stars.

A detailed description of the whole error estimate can be found  e.g.in Paper IV
and will not be repeated here, as it is beyond the aim of the present discussion.
 The interested reader may find an extensive discussion in the appendix  
of the present paper, with tables of
sensitivities, estimates of the actual errors in the atmospheric parameters, and
resulting uncertainties in abundances. In Appendix A we clearly separate the 
individual, star-to-star errors (relevant to the discussion of the abundance
spread in each cluster) from the cluster errors, which concern the whole cluster
sample.

The expected star-to-star scatter in [Fe/H] caused by the three major  ($T_{\rm
eff}$, $v_t$, $EW$) or to all error sources (last two columns in Table A3 in the
Appendix) may be compared to the $observed$ scatter (defined as the $rms$
scatter of all stars in each cluster, column 5 in Table~\ref{t:quantefe}).  We
note that, for at least half of our sample, the expected scatter is formally
higher than the observed spread, even taking the statistical
uncertainty into account. This may be due to an overestimate of some error sources
or to correlations  and it does not invalidate the conclusion that globular
clusters are very homogeneous (concerning Fe content) objects.
 Most of our programme clusters are homogeneous in [Fe/H] at a level
below 10\%, and when higher quality data are available (as in NGC~5904, NGC~6121)
the level that any theoretical model of cluster formation has to reproduce drops
to a 6\% degree of homogeneity involving products from supernovae
nucleosynthesis. We stress that this is a very strong constraint to be
satisfied.

Finally, typical star-to-star errors are 0.14 dex in [O/Fe] and 0.08 dex in
[Na/Fe], on average (see Appendix A, Table A2).

\section{Results and discussion}

\begin{figure}
\centering
\includegraphics[scale=0.45]{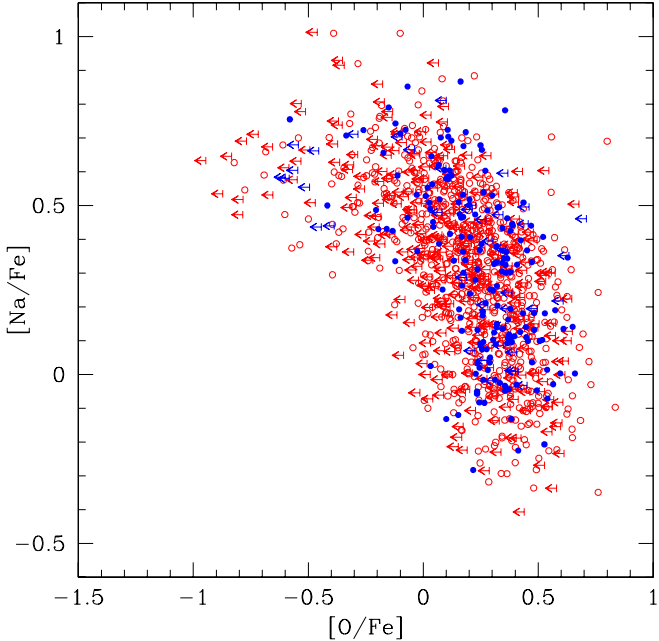}
\caption{The Na-O anticorrelation for a grand total of 1958 individual red
giant  stars in the 19 GCs of our project. [Na/Fe] and [O/Fe]
ratios from GIRAFFE spectra are shown as open (red) circles; abundance ratios
obtained from UVES spectra (Paper VIII) are superimposed as filled (blue)
circles and show no offset from the GIRAFFE sample. Arrows indicate upper
limits in oxygen abundances.}
\label{f:naaah}
\end{figure}

\subsection{The Na-O anticorrelation}

We derived abundances of O and Na from measured $EW$s. In principle, among the
1409 member stars observed with FLAMES/GIRAFFE and with atmospheric parameters 
and Fe determination in our 15 clusters we could expect to measure O in a
maximum of 1089 stars, all those observed with the HR13 grating.

However, although all oxygen lines were carefully inspected by eye, the
combination of unfavourable observational constraints (too low a S/N,
the faintness of stars in not well-populated clusters) and/or of physical
ingredients (very large O-depletions, cluster low metallicity) prevented the O
abundance to be derived in all stars. We measured O abundances in a subsample of
865 stars, including 313 upper limits.

Oxygen abundances were obtained from the forbidden  [O {\sc i}] lines at 6300.3
and 6363.8~\AA;  the former was cleaned from telluric contamination by
H$_2$O and O$_2$ lines using a synthetic spectrum, as described in Paper I. Our
experience with the analysis of the first four clusters is that the contribution
of the high excitation Ni {\sc i} line at 6300.34~\AA\ to the measured $EW$ is
negligible (see also Paper II), and the CO formation does not have a relevant impact
on the derived O abundances due to the rather high temperature of our programme
stars.

\begin{table*}
\centering
\caption{Number of stars with both Na and O and fraction of the primordial,
intermediate, and extreme components}
\begin{tabular}{lrrrrrr}
\hline
GC       &N.stars & N.stars & N.stars&fraction &fraction &fraction \\
         &(O,Na)  & (O,Na)  & (O,Na) &P        &I	 &E	   \\
         &GIRAFFE &GIR.+corr&GIR.+UVES   &component&component&component \\
\hline
NGC  104 &109 & 109 & 115&  $27\pm ~5$ &  $69\pm ~8$ &  $ 4\pm 2$  \\
NGC  288 & 64 &  64 &  70&  $33\pm ~7$ &  $61\pm ~9$ &  $ 6\pm 3$  \\
NGC 1904 & 49 &  39 &  48&  $40\pm ~9$ &  $50\pm 10$ &  $10\pm 5$  \\
NGC 2808 & 91 &  91 &  98&  $50\pm ~7$ &  $32\pm ~6$ &  $18\pm 4$  \\
NGC 3201 &104 &  94 & 100&  $35\pm ~6$ &  $56\pm ~7$ &  $ 9\pm 3$  \\
NGC 4590 & 48 &  36 &  44&  $40\pm ~9$ &  $60\pm 11$ &  $ 0^{+4.1}_{-0.0}$  \\
NGC 5272 &    &     &  37&  $32\pm ~9$ &  $68\pm 14$ &  $ 0^{+4.6}_{-0.0}$  \\
NGC 5904 &106 & 106 & 114&  $27\pm ~5$ &  $66\pm ~8$ &  $ 7\pm 2$  \\
NGC 6121 & 80 &  80 &  88&  $30\pm ~6$ &  $70\pm ~9$ &  $ 0^{+2.1}_{-0.0}$  \\
NGC 6171 & 27 &  27 &  30&  $33\pm 11$ &  $60\pm 14$ &  $ 7\pm 5$  \\
NGC 6205 &    &     &  53&  $34\pm ~8$ &  $45\pm ~9$ &  $21\pm 6$  \\
NGC 6218 & 67 &  67 &  74&  $24\pm ~6$ &  $73\pm 10$ &  $ 3\pm 2$  \\
NGC 6254 & 99 &  77 &  87&  $38\pm ~7$ &  $60\pm ~8$ &  $ 2\pm 2$  \\
NGC 6388 & 29 &  29 &  32&  $41\pm 11$ &  $41\pm 11$ &  $19\pm 8$  \\
NGC 6397 &  6 &   3 &  16&  $25\pm 13$ &  $75\pm 22$ &  $ 0^{+12.0}_{-0.0}$  \\
NGC 6441 & 24 &  24 &  29&  $38\pm 11$ &  $48\pm 13$ &  $14\pm 7$  \\
NGC 6752 & 89 &  89 &  98&  $27\pm ~5$ &  $71\pm ~9$ &  $ 2\pm 1$  \\
NGC 6809 &105 &  75 &  84&  $20\pm ~5$ &  $77\pm 10$ &  $ 2\pm 2$  \\
NGC 6838 & 31 &  31 &  42&  $29\pm ~8$ &  $71\pm 13$ &  $ 0^{+4.2}_{-0.0}$  \\
NGC 7078 & 37 &  20 &  33&  $39\pm 11$ &  $61\pm 14$ &  $ 0^{+5.5}_{-0.0}$  \\
NGC 7099 & 27 &  19 &  29&  $41\pm 12$ &  $55\pm 14$ &  $ 3\pm 3$  \\

\hline
\end{tabular}
\label{t:quantenao}
\end{table*}

Sodium abundances could be obtained for many more stars, since at least one
of the Na~{\sc i} doublets at 5672-88~\AA\ and at 6154-60~\AA\ is always 
available (depending on the GIRAFFE setup used). Again, the Na measurements were
interactively checked by eye in all cases where clear discrepancies between
abundances from the 2 to 4 different lines were present. Derived average Na
abundances were corrected for effects of departures from the LTE assumption
according to the prescriptions by Gratton et al. (1999).

This was our first step and it produced the number of stars with both O and Na
abundances derived from GIRAFFE spectra listed in column 2 of
Table~\ref{t:quantenao}, where for completeness we included also the number of 
stars used in the Na-O anticorrelation in the four previously analysed clusters.

Afterward, we checked for possible systematic effects in Na abundances as
derived from the two doublets. On average, there are no large systematic
differences, the mean difference in the sense 6154-60~\AA\ minus 5682-88~\AA\ 
being $\Delta \log{\rm n(Na)} = +0.001\pm 0.007$ dex, with $rms=0.181$ dex from
678 stars. 

However, we studied a large sample of stars in clusters spanning almost 2 dex 
in metallicity, and we detected a subtle statistical bias 
 by plotting the differences as a function of [Fe/H]. When the Na~{\sc i} lines at 6154-60~\AA\
are very weak, they are measurable only when spuriously enhanced by noise. This
suggests that we can overestimate the Na abundance using these lines  in
particular in metal-poor and warmer stars. To correct for this effect we used an
empirical parameter, defined as  ($T_{\rm eff}/100$)$-10 \times$[Fe/H].

If this parameter was larger than 65, then
\begin{itemize} 
\item if only 
lines belonging to the 6154-60~\AA\ doublet were available for the star, they were
eliminated and the star was thus dropped from the Na-O anticorrelation;

\item for stars with 2, 3, 4 lines of Na, average [Na/Fe]$>0.2$ dex and
$rms(Na)<0.2$ dex, all the lines were retained;

\item for stars with 2, 3, 4 lines
of Na and $rms(Na)>0.2$ dex, the 6154-60~\AA\ lines were deleted;
\end{itemize}

After this correction (culling out stars, in particular in the most metal-poor 
clusters), the number of stars participating to the Na-O anticorrelation is the
one listed in column 3 of Table~\ref{t:quantenao}.

Finally, our third step was to combine chemical composition measurements
derived from the 
GIRAFFE spectra sample with
Na and O abundances derived from the analysis of UVES spectra, for which 
analysis and element ratios are discussed in Paper VIII. Regarding Fe, Na, and O,
it suffices to say here that we followed the same procedures used for the
GIRAFFE spectra, both to obtain atmospheric  parameters and the abundance
ratios.

\begin{figure*}
\centering
\includegraphics[bb=20 150 580 600,clip, scale=0.89]{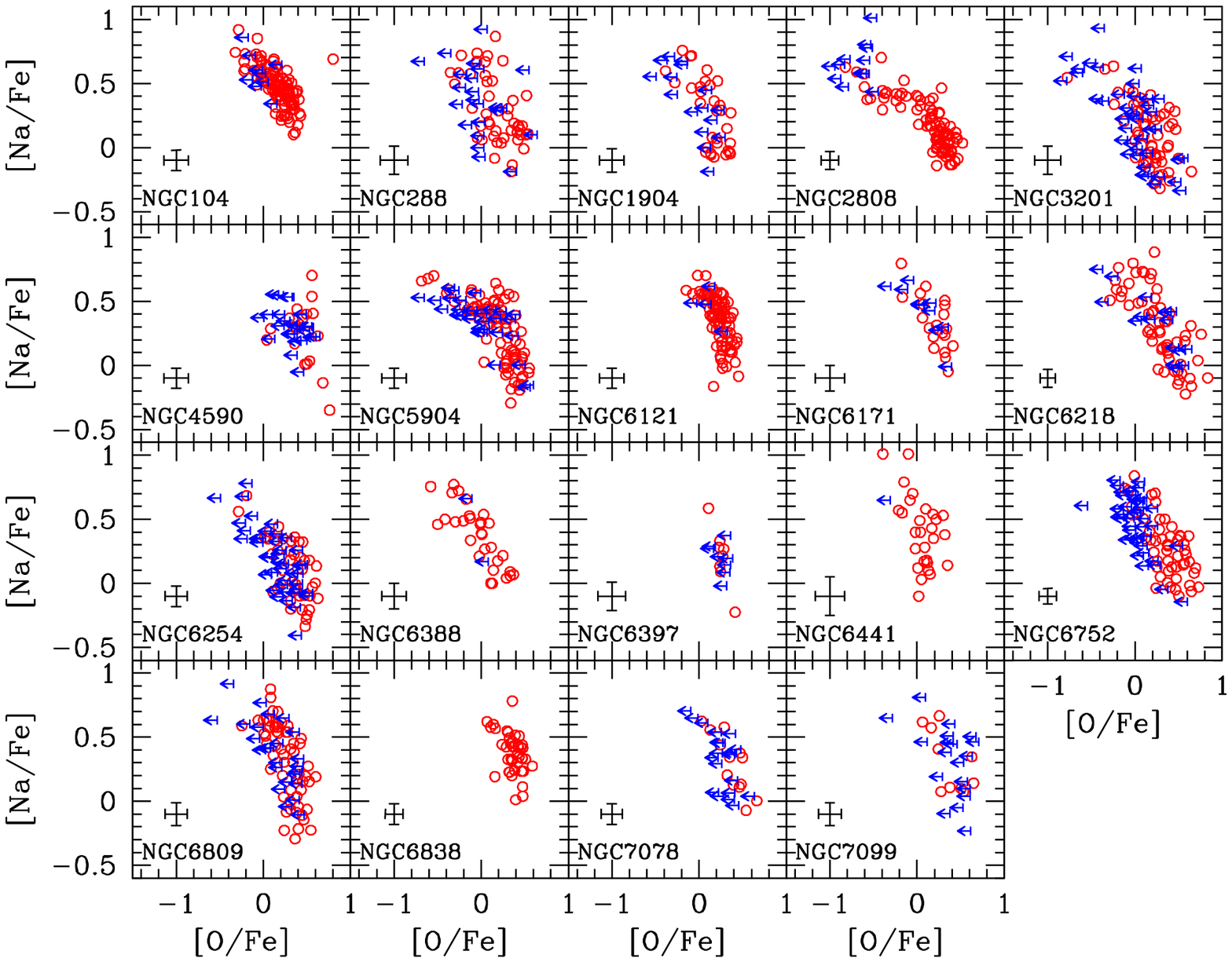}
\caption{The Na-O anticorrelation observed in all the 19 GCs of
our project. All stars with Na and O abundances from GIRAFFE and UVES (Paper
VIII) spectra are used. Star-to-star error bars (see Appendix A) are indicated
in each panel. Upper limits in O abundances are shown as arrows, detections are
indicated as open circles.}
\label{f:tutteantino}
\end{figure*}

There are 214 stars with UVES spectra analysed in the 19 clusters of our
complete sample; of these, 172 stars are in the 15 clusters of the present work, 170 of
which have both O and Na. [Na/Fe] and [O/Fe] abundance ratios from UVES spectra
are superimposed to the same ratios from GIRAFFE spectra in Figure~\ref{f:naaah}. 
This figure shows that there is no obvious offset between the two data sets 
and, together with the very good agreement obtained in iron abundances (see
Figure~\ref{f:FeI}), this guarantees that the two samples can be safely merged
without introducing any bias.

This is a crucial point for some clusters, especially for NGC~6397, where only
a handful of O detections (mostly upper limits) could be extracted from the
GIRAFFE spectra. Hence, the final step in exploring the Na-O anticorrelation
in our programme clusters was to substitute O and Na values obtained from the UVES
spectra for stars observed with both instruments and to add the values from
stars with only UVES observations.

In Table~\ref{t:abunao} we list the abundances of O and Na (the complete table
is available only in electronic form at CDS) in each star of the present
subsample of 15 GCs. For O we distinguish between  actual
detections and upper limits. The number of measured lines and the rms values are
also indicated.

Column 4 of Table~\ref{t:quantenao} provides the final numbers of stars that we
used to build the Na-O anticorrelation in each of the 19 clusters of this
project. We have a grand total of 1235 red giants with O and Na abundances
derived homogeneously (936 in the 15 clusters analysed here),  by far {\it the 
largest sample collected up to date}.

\begin{table*}
\centering
\caption[]{Abundances of O and Na for the 1409 stars with only GIRAFFE spectra  
in 15 GCs. The complete Table is available only in electronic form at CDS.}
\begin{tabular}{rrcccccccc}
\hline
GC & Star     & nr &[O/Fe]& rms  &  nr &[Na/Fe] &rms &HR & lim \\    
\hline                                                           
 NGC~104 &  1389 & 2  &  0.395& 0.069	& 2  & +0.175 &0.011 & 1 & 1\\ 
 NGC~104 &  2608 & 1  & -0.207& 	& 4  & +0.615 &0.078 & 2 & 1\\ 
 NGC~104 &  2871 &    &       & 	& 2  & +0.440 &0.062 & 3 & 1\\ 
 NGC~104 &  4373 & 1  &  0.430& 	& 4  & +0.249 &0.135 & 2 & 1\\ 
 NGC~104 &  5172 & 2  &  0.189& 0.002	& 4  & +0.489 &0.072 & 2 & 1\\ 
\hline
\end{tabular}
\label{t:abunao}
\end{table*}

In Figure~\ref{f:tutteantino} the Na-O anticorrelation we obtain in all the 19
clusters is shown, with star-to-star error bars plotted in each panel. In these
plots we used all available stars in each cluster with both Na and O abundances,
irrespective of their derivation from GIRAFFE or UVES (Paper VIII) spectra.

We also searched the literature for GCs not included
in our programmes, with a large ($>30-40$) number of stars analysed, and with O
and Na  abundances from high-resolution spectra. We only found two GCs meeting
these requirements: NGC~5272 (M~3) and NGC~6205 (M~13). For these clusters we
used the stars analysed in the most recent studies (Sneden et al. 2004 and Cohen
and Melendez 2005), corrected to our scale of solar reference abundances, and
merged their samples with ours, adopting for stars in common those from Sneden et al.
The final adopted numbers of stars are reported in column 4 of
Table~\ref{t:quantenao}.

For several clusters in our sample, this is the first-ever survey of this kind
based on a very large numbers of stars. 
For example, since it is a nearby and luminous cluster, 47~Tuc is
often used as a yardstick for abundance analysis, but only a few 
stars were previously observed and analysed. To our knowledge, our homogeneous
database of 115 red giants in this cluster is the largest collected to extensively
study the Na-O signature in this object. Within the present project, the Na-O
anticorrelation is traced and also studied for the first time for several
other clusters: NGC~1904, NGC~2808 (apart from 19 stars from Carretta et al.
2004b), NGC~4590, NGC~6171, NGC~6397, NGC~6441, NGC~6809, and NGC~7099. The wide
range in HB morphology, metallicity and other cluster parameters strengthens the
suggestion (see Carretta 2006) that this signature is present in $all$ clusters
where data allows us to investigate it, and it is probably related to the same mechanism of
formation and early evolution of GCs.

In some cases the number of stars available to probe the Na-O anticorrelation is
limited by the number of stars that turned out to be actual cluster members. This
was the case for the disc clusters NGC~6171 and NGC~6838 and for the bulge
clusters NGC~6441 (Paper V) and NGC~6388. In very metal-poor clusters the number
of stars in the [Na/Fe]--[O/Fe] plane is lower than expected because of the
difficult task of measuring in particular the forbidden [O~{\sc i}] lines, the
worst case being NGC~6397 where only the addition of measurements from UVES
spectra allows us to derive the observed anticorrelation.

The distribution function of the [O/Na] ratios from our data (including both
GIRAFFE and UVES observations) is shown in Figure~\ref{f:histoonan} for all the 15
clusters analysed here plus the four clusters already studied. In each panel,
the histograms are normalised to the total number of stars with O and Na
abundances.

\subsection{The primordial, intermediate, and extreme components}

The presence of large star-to-star variations in abundance of elements that 
cannot be produced in presently observed low-mass red giants is the clearcut
proof of the existence in GCs of at least two different stellar
generations.

The ratio of the number of first to second-generation stars could be very useful for
constraining any formation scenario (see e.g., D'Ercole et al. 2008). However,
to truly be meaningful, such a quantity must be derived from large samples of
stars all analysed in the same way, to avoid introducing spurious effects 
reflecting possible offsets in the analyses.

Our database offers the unique, unprecedented opportunity to study the behaviour of about 1,600
red giants in a significant fraction of the whole galactic GC
population. We sampled red giant stars with no obvious bias with respect to
their Na and O abundances. We could not measure O abundances in
all stars, and we only placed upper limits to O abundances in many stars, generally
warm, metal-poor, and O-poor. In spite of this limitation, we think that our
sample allows a statistically robust estimate of the fraction of stars formed in
different bursts within GCs,  with a caveat about this selection effect.

We assume the first-generation (or P) to be those stars
with O and
Na content similar to field stars of the same  metallicity [Fe/H]. The latter are
usually characterised by a pattern typical of supernova nucleosynthesis with
quite uniform super-solar O values and slightly sub-solar  Na abundances, the
exact value depending on metallicity (with some scatter).  Hence, in each cluster,
we assigned stars to the P component if their [Na/Fe] ratios fall in the range
within [Na/Fe]$_{\rm min}$ and [Na/Fe]$_{\rm min} + 0.3$ (that is $\sim
4\sigma{\rm ([Na/Fe])}$, where  $\sigma{\rm ([Na/Fe])}$ is the star-to-star
error on [Na/Fe] in each cluster. The minimum value for the ratio [Na/Fe] in
each cluster was estimated  by eye by looking at the anti correlations in
Figure~\ref{f:tutteantino}, excluding obvious outliers. They are listed in
Table~\ref{t:dilution} and  match the [Na/Fe] ratios observed in
field metal-poor stars  quite well (see Section 7). With this criterion we are confident that
we have included all the primordial stars, i.e. those with typical composition of normal
halo stars, although a few stars with slightly modified abundances might be included,
too, so this definition may somewhat overestimate the P population.

\begin{figure}
\centering
\includegraphics[bb=40 160 610 700, clip, scale=0.46]{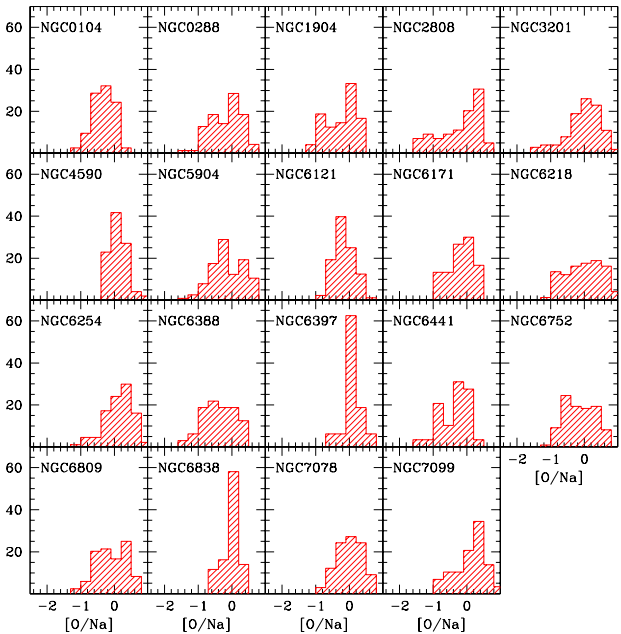}
\caption{Distribution function of the [O/Na] ratios along the  Na-O
anticorrelation in all the 19 programme clusters of this project. The histograms
are normalised to the total number of stars used in each cluster.}
\label{f:histoonan}
\end{figure}

\begin{figure}
\centering
\includegraphics[scale=0.44]{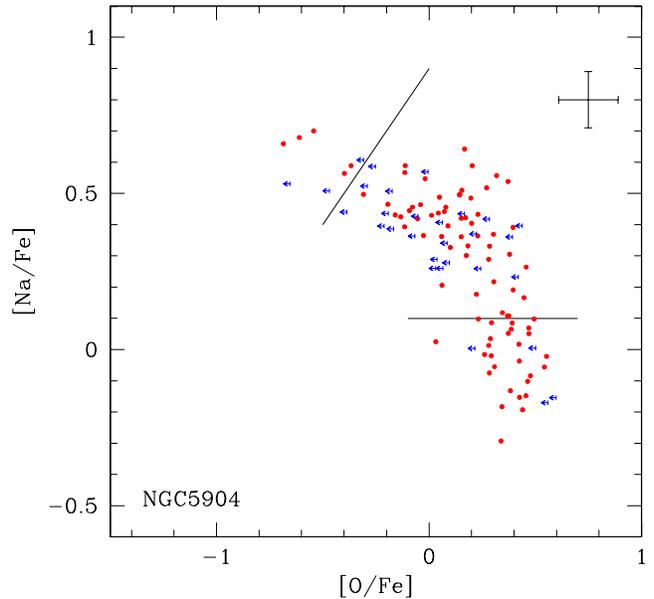}
\caption{The Na-O anticorrelation from our data observed in NGC~5904. The solid
lines indicate the separations we adopted for the P, I, and E 
stellar components in this cluster.}
\label{f:pie}
\end{figure}

The remaining stars departing from this high-O, low-Na locus along the
anticorrelation are considered all second-generation stars. We further  divided
this group by how much the abundances depart from those of the P
population: stars with the ratio [O/Na]$>-0.9$ dex are assigned to an
intermediate (I) component, while those with [O/Na]$<-0.9$ dex belong to the
extreme (E) stellar component of second-generation cluster stars. We chose this
separation by comparing the distribution functions of the [O/Na] ratios in all
clusters  (see Figure~\ref{f:histoonan}). This limit is arbitrary and corresponds
to a minimum or a sudden drop in the [O/Na] distribution clearly discernible in
the distribution of some clusters (NGC~2808, NGC~5904, NGC~3201), where a long
tail of O-depleted stars was reliably measured. In Figure~\ref{f:pie} the lines
separating the three components are shown using NGC~5904 as an example.

We applied these criteria to all 19 our programme clusters and to the two clusters
from the literature. Only the separation between the first and the
second-generation stars changes, since it is tied to the minimum Na abundances
that, as in field-halo stars, include a slight dependence on the metallicity.
The fractions of stars in the three P, I, and E components in each cluster are
listed in the last three columns of Table~\ref{t:quantenao}. Associated errors
are computed from Poisson's statistics. In cases where no stars were found in a
group (i.e., the E population), we evaluate the errors as  the probability of
occurrence of zero stars to be retrieved in a sample of stars (equal to the
total number of stars in the anticorrelation) according to the  binomial
distribution.

\begin{figure}
\centering
\includegraphics[scale=0.45]{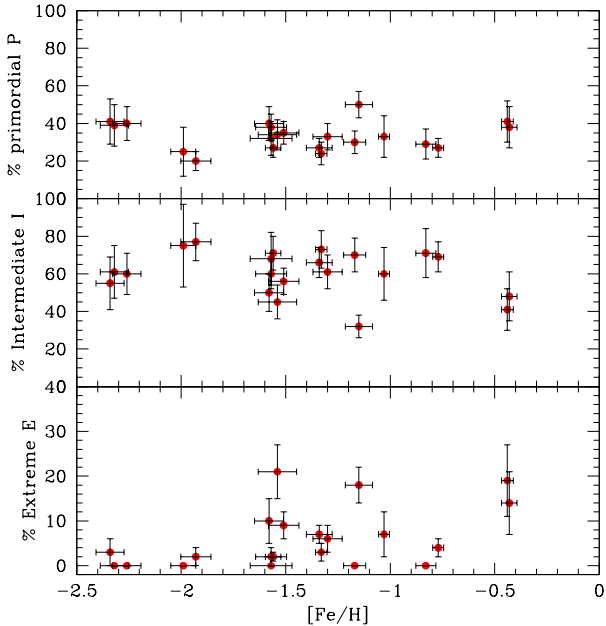}
\caption{Fractions of stars in the P, I, and E
stellar components (upper, middle, and lower panels respectively) derived from
the Na-O anticorrelation in our 19 clusters and in M~3 (NGC~5272) and M~13
(NGC~6205) from Sneden et
al. (2004) and Cohen and Melendez (2005) as a function of the metallicity. Error
bars in the fractions are estimated Poisson's statistics. For metallicity we used the
cluster errors (Table~\ref{t:quantefe}) for our sample and the quadratic sum of
the $rms$ scatters from the two studies for M~3 and M~13. Notice that the scale
of y-axis is different in the lower panel.}
\label{f:piefe}
\end{figure}

These fractions are plotted as a function of metallicity in Figure~\ref{f:piefe},
where we used cluster errors from Table~\ref{t:quantefe} for our sample; for 
M~3 (NGC~5272)
and M~13 (NGC~6205) error bars in [Fe/H] are the quadratic sum of the $rms$ scatters quoted
in Sneden et al. (2004) and Cohen and Melendez (2005), since no systematic
errors are derived in the original studies. 

From the upper panel in Figure~\ref{f:piefe} it is immediately clear that a
P component, which can be identified as the original, first-generation
of stars, seems to be present at a constant level of about one third of the 
total population in $all$ clusters surveyed. The average fraction we found for 
the set of 21 clusters is P$=33\pm1\%$ with $rms=7\%$ over the whole 2 dex range
in metal abundance.

How statistically robust is this estimate? The three components are defined 
using stars in the [Na/Fe] vs [O/Fe] plane. However, the criterion for the P 
component only uses the Na abundances; hence, for this component only, we may 
explore the impact of adding those stars with Na but without O abundances. 
In our total database there are 511 objects with only Na determinations; 377
stars have no HR13 observations, the others are all quite warm ($T_{\rm
eff}$\ between 4600 and 5400 K), and metal-poor stars where the forbidden O lines
can be vanishingly weak even in stars in the high-O, low-Na tail of the
anticorrelation. 
On the other hand, Na abundances can be recovered more easily
since (i) we can exploit the stronger 5682-88~\AA\ Na~I doublet for the majority
of stars and (ii) the Na-depletion at this extreme is not as much as the
O-depletion at the opposite end of the Na-O anticorrelation.  Using this
additional set of 511 stars, we computed again the fraction of the P component
in our sample. Despite the increase in statistics, the new values of the P
fractions changed on average by $0\pm1\%$ ($rms=5\%$) for 19 clusters, but the
change never exceeded 8-9\%. In addition, the main statistical bias  present in
our data (the upper limits for O abundances in many stars) does not affect this
parameter, which is only based on Na abundances. Only for the most metal-poor
clusters (like M15) might we have missed the most Na-poor stars, producing some
bias. In these cases Na$_{\rm min}$\ might have been overestimated; however, the
impact on the fraction on stars in the P population is small. So, we can
consider the estimate of the P (first  stellar
generation) fraction  in GCs as quite robust.

The fraction of the stars belonging to the I component
(middle panel of Figure~\ref{f:piefe}) would also seem about constant (at a level
$\sim 65\%$) except for three clusters (NGC~2808, NGC~6388, and NGC~6441, all
massive and with long blue tails on the HB)  where this fraction is clearly
smaller.

Finally, the fraction of component E shows the largest fluctuations, being null or
very low in many clusters, raising to a modest 10\% in a few, and increasing  to
about 20\% in 4 clusters. Three of them are the objects with a lower-than-average
I component seen above; to these, we can add M~13 (NGC~6205) showing an almost normal I
component, but a conspicuous E stellar fraction.

On the other hand, it is not easy to assess how stable our estimates are
concerning the two second-generation components (I and E), since by definition
we need to know $both$ Na and O abundances to assign a star to one group or to
the other. Since we derived only upper limits for O abundances for a significant
fraction of the stars, we might have underestimated the fractions of stars
belonging to the E component; hence, should the E fraction be larger, in
reality, the complementary I fraction would be smaller, by definition.

In some clusters we are quite confident that the E fraction cannot be much
higher than estimated: the higher quality of data and the metallicity 
for 47~Tuc (NGC~104)
or M~4 (NGC~6121) result in very few limits, most O determinations being actual measures.
Second-generation stars with E chemical composition are simply missing in
these clusters (for M~4 this is strongly supported by the recent study by 
Marino et al. 2008). In other cases, such as in NGC~6752, where our data are of
poorer  quality and we only got upper limits to O abundances for quite a large
fraction of the stars, the high-resolution/high S/N data by Yong et
al. (2005) show that our upper limits in O can be safely considered as actual
measures and that very few or no super O-poor  stars of the E component might
be expected to show up in this cluster (see the discussion in Paper II).

Thus, the first conclusions we can draw from our data can be summarised as 
follows:
\begin{itemize}
\item a P population is present in all GCs; about a third of the
cluster population is still made of the original first-generation, after a
Hubble time since the cluster formation;
\item the I component of the second-generation constitutes the bulk 
(50-70\%) of stars in the clusters;
\item E, the second-generation component with signature of extreme chemical 
composition is not present in all GCs.
\end{itemize}

\subsection{The radial distribution of first and second-generation stars}

In the Introduction, we recalled the strong existing pieces of evidence indicating that
the Na-O anticorrelation is related to multiple populations in GCs. 
The pattern of chemical composition is the result of stellar 
nucleosynthesis and ejection of polluted matter. The distribution of stars along
the Na-O anticorrelation may be reproduced by diluting the polluted material
with pristine gas before second-generation stars form (see Prantzos,  Charbonnel
\& Iliadis 2007). However, we still do not know whether the polluters  of the
first-generation contributed their enriched matter to the intra-cluster  pool of
gas in their main sequence phase (as fast-rotating massive stars) or  in a more
evolved stage (as massive AGB stars): see Decressin et al.  (2007), D'Antona \&
Ventura (2007), Renzini (2008). However, we expect that second-generation stars
should be He-rich.

Second-generation stars might be expected to form (and perhaps still
be) more centrally concentrated than first-generation stars (see
D'Ercole et al. 2008).  In fact, the spatial distribution of first-generation 
stars is expected to be loose because of the cluster
expansion from the large amount of mass lost by massive stars in the
very early phases of cluster evolution. On the other hand, we could
expect that later stellar-generations form from a cooling flow at the
cluster centre and are (at least initially) kinematically very cold.
These different distributions should result in very different rates of
evaporation, first-generation stars being lost by the cluster much
more easily than second-generation ones during the early epochs of
cluster evolution. On the other hand, stars with He-enhanced
composition evolving off the main sequence are expected to be
(slightly) less massive than those with ``normal'' composition
(D'Antona et al.  2002). In the long dynamical evolutionary phase
dominated by the two-body relaxation, the cluster is driven toward
equipartition of kinetic energy. It is then possible that, after a
Hubble time (and several relaxation times), He-enhanced (O-poor,
Na-rich) red giants might have a more extended distribution than
He-poor ones.

Very recently, Zoccali et al. (2009) have found that the peculiar second
subgiant branch observed in NGC~1851 is only present in the central
regions of the cluster, disappearing at about 2.4 arcmin from the
cluster centre, and it is well known that the blue, He-enriched main
sequence in $\omega$ Cen is more centrally concentrated than the
He-normal sequence (Sollima et al. 2007).

We can test the spatial distribution of stars in the first and second
generations using our database, keeping in mind the practical
limitations imposed by the FPOSS positioner to the FLAMES fibres.

\begin{figure}
\centering
\includegraphics[scale=0.45]{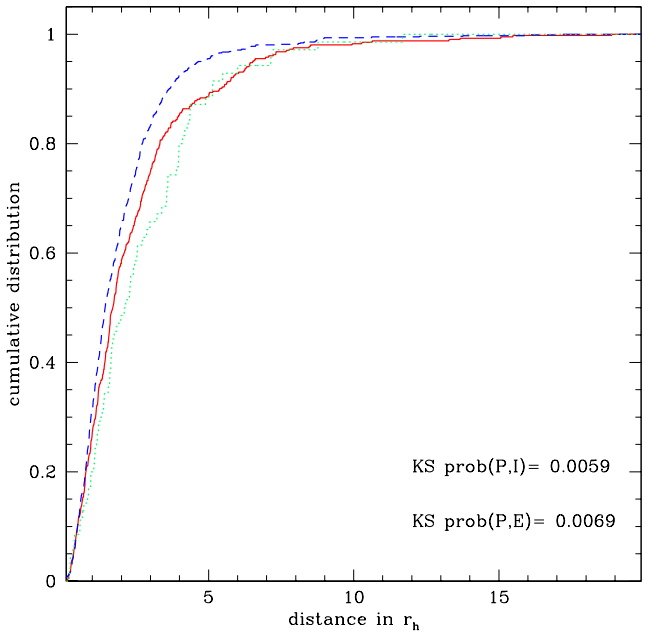}
\caption{Cumulative distribution of stars of the 3 components in our 19 clusters
(plus M~3=NGC~5272 and M~13=NGC~6205)
in unit of half-mass radii. Red solid line: component P, blue dashed
line: component I, green dotted line: component E.}
\label{f:radialecumul}
\end{figure}

The cumulative radial distributions of stars in the three P, I, E
components are shown in Figure~\ref{f:radialecumul}, including the
two additional clusters M~3 (NGC~5272) and M~13 (NGC~6205). Apparently, despite  being 
forced to observe at some distance from the centre of the GCs (to
maximise the number of targets in each cluster, while avoiding
forbidden positions of the fibres), this figure shows that the
I component is more concentrated than the P
component. A Kolmogorov-Smirnov statistical test excludes that the two
distribution are extracted from the same parent population, with only
a 0.6\% probability that this is a chance occurrence, so we remind the reader that
these are the two most conspicuous components in each cluster. From
the same figure it is unclear how much the E component is differently
distributed with respect to the P component; from the
Kolmogorov-Smirnov test, the probability that they are extracted only
by chance from a same parent population is 0.7\%.

\begin{figure}
\centering
\includegraphics[scale=0.45]{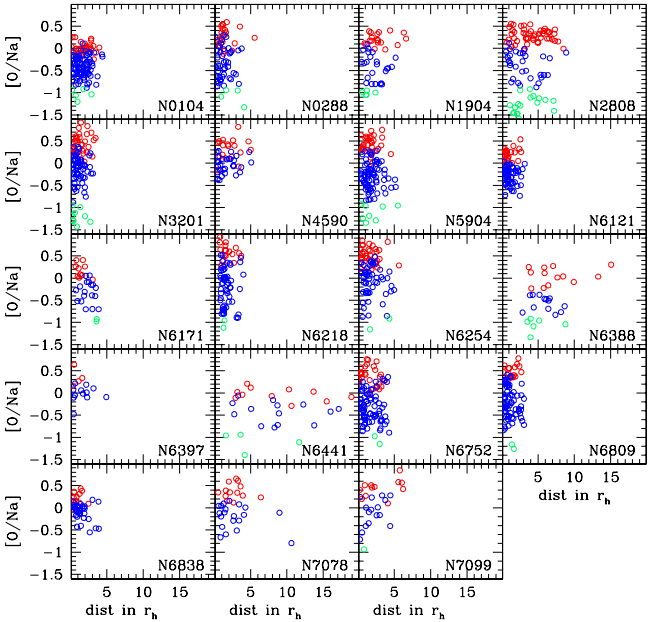}
\caption{[O/Na] ratios for all stars observed in our 19 programme clusters as a
function of the distance from the cluster centre, expressed in units of the
half-mass radius. Red, blue, and green symbols are for P, I, and E populations,
respectively.}
\label{f:radialerh}
\end{figure}

However, there is the possibility that the cumulative distributions in
Figure~\ref{f:radialecumul} are biased. In fact, although distances in
different clusters are all expressed in units of half-mass radius (the
region where most cluster properties are left relatively unchanged by
the evolution), our programme clusters have different masses, sizes, and
central concentrations; hence, we could have observed regions that
are not dynamically equivalent in all clusters. This is evident in
Figure~\ref{f:radialerh}, where we plotted the [O/Na] ratios for stars
in each cluster as a function of the distance (in $r_h$ units).

To check this effect we proceeded as follows. In each cluster, we
computed the median of the distances from the cluster centre for each
component distP$_{med}$, distI$_{med}$, and distE$_{med}$. Each median
was normalised to that of the I population, which is the
most numerous in each cluster.  Afterward, we computed the average of the
normalised medians for the P and E populations, and these averages are
$<$distP$_{med}$(normalised)$>$ = 1.329 with $\sigma=1.292$ and
$<$distE$_{med}$(normalised)$>$ = 1.151 with $\sigma=0.765$ from 21
and 15 clusters, respectively\footnote{Obviously,
  $<$distI$_{med}$(normalised)$>$ = 1.0 by definition.}. Although
formally this might indicate that the P stars are more externally
distributed, on average, than the I ones (and the E still more),
the difference is not significant.  The large scatter relative to the
first average is all due to the value for M~3 (NGC~5272). This cluster was
observed very near to the centre, because of the requirement of putting as
many RGB stars as possible in the observing masks (see Sneden et al.
2004 for details). The impression is that differences in the spatial
distributions of stars in the three components might exist, but they
are somewhat smeared out by the bias from observing different
dynamical regions in the GCs.

\begin{figure*}
\centering
\includegraphics[bb=19 163 586 490, clip, scale=0.69]{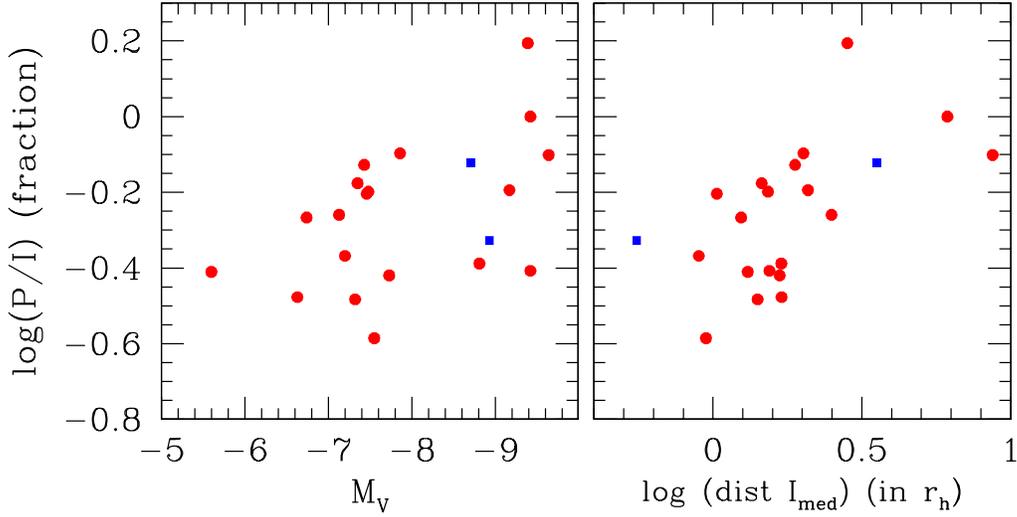}
\caption{The logarithm of the ratio between the fraction of stars in the
P and the I components in each programme cluster as
a function of the cluster total absolute visual magnitude (from Harris 1996),
in  the left panel, and of the median of the distances of stars in the I
component from the cluster centre (in unit of half-mass radius), in the
right panel. Red filled circles are our programme clusters and blue squares
indicate the two additional GCs from the literature.}
\label{f:percdistmv}
\end{figure*}

This impression is strengthened by Figure~\ref{f:percdistmv}, where we
plot the ratio of the fraction of P to I component
as a function of the absolute visual magnitude (a proxy for the
cluster mass) in the left panel and as a function of the median
distance of the I component (a proxy for the typical position at which
we observe the cluster, since the I stars are the bulk of the
clusters' population) in the right panel. From this figure we can see
(left panel) that, by looking at more massive clusters, we observe a
larger fraction of P stars w.r.t. the I component.
This would have a simple physical explanation because it is expected
that massive objects are able to retain a larger fraction of their
stars, including their first-generation stars. However, the right panel
of Figure~\ref{f:percdistmv} shows that the P fraction is also larger in
clusters where we typically sampled more peripheral regions in the GC.
The same holds had we used the ratio of P to the sum of I and E, i.e.
the ratio of first to second-generation stars, without separating the
two I and E components.

We can evaluate the order of magnitude of this effect by computing a
``corrected'' P/I ratio from the right panel of
Figure~\ref{f:percdistmv}. Although the scatter in this plot is quite
large, we can fit a straight line and thus get the value of
$\log$(P/I)$_{corr}$ that takes the position into account at which the
cluster was observed (as expressed by the median of the distances of
stars in the I component, in units of half-mass radius). By applying
this correction we find that the ratio of P to I
stars, when shifted to a reference half-mass ratio, is about
constant (-0.26 in logarithm). In other words, had we always observed
the bulk of our programme stars at the cluster half-mass ratio, we would
have found that the P component is about 55\% of the
I one. Using the ratio of first to second-generation stars
(the last including both I and E components) we would have found that
on average from 47 to 49\% of stars in clusters are from the pristine
stellar generation formed in each cluster.

This exercise, while clarifying some operative issues, does
not, however, solve that related to the true distribution of the three
components across a GC.  More observations of larger
samples of stars in the smaller clusters will be needed to definitively
solve the issue of the radial distribution of stars of different
generations in GCs.

\section{Nitrogen abundances of first and second-generation stars}

The whole pattern of inter-relations among light elements in globular
clusters is currently well known (see e.g. the review by Gratton et
al. 2004). However, up to now, these signatures have been poorly
explored with respect to the membership of stars to one stellar
generation or another in a GC. The recent paper by Marino et al.
(2008) found that the dichotomy in chemistry (mainly in O, Na and N
content) between two generic populations in NGC~6121 was also visible
as a different photometric location along the RGB of the two groups.
Using the $U-B$ colour, strongly affected by N abundances due to the
location of NH (around 3360~\AA) and CN (at about 3590 and 3883~\AA)
features, they clearly showed that Na-poor/O-rich/N-poor stars define
a sequence to the blue ridge of the RGB, whereas Na-rich/O-poor/N-rich
stars are more spread out, to the red of the RGB.

\begin{figure*}
\centering
\includegraphics[scale=0.30]{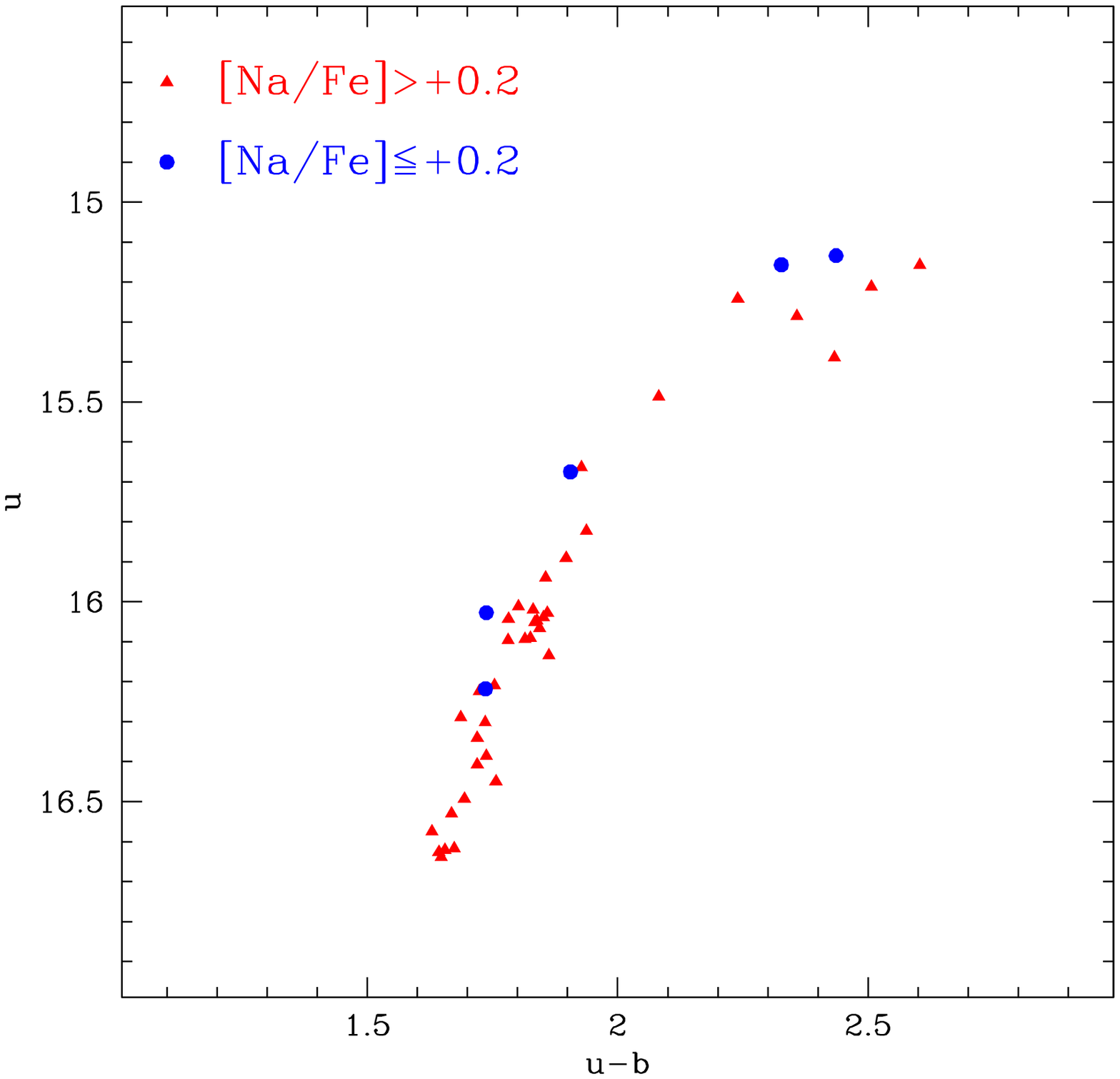}\includegraphics[scale=0.30]{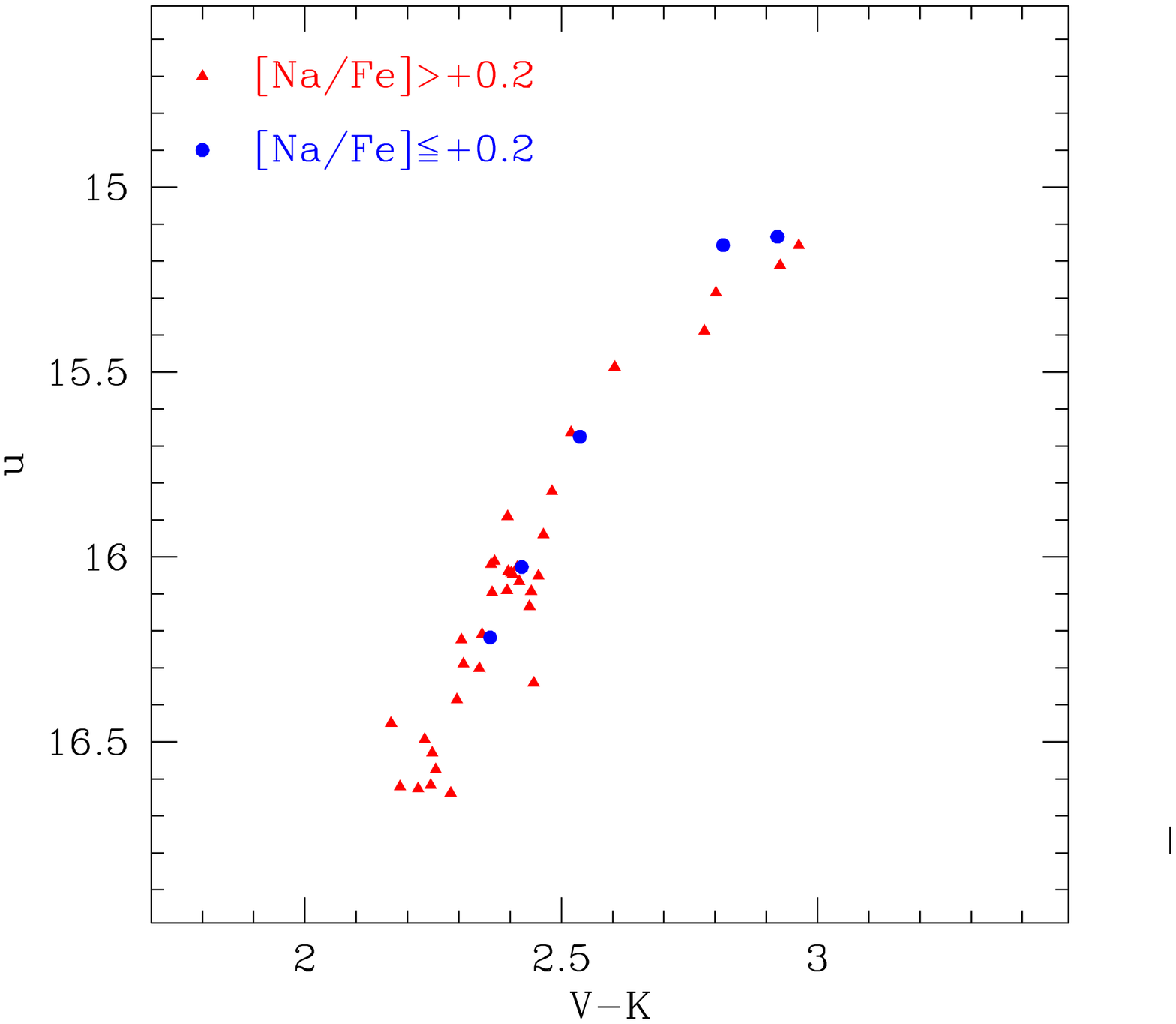}\includegraphics[scale=0.30]{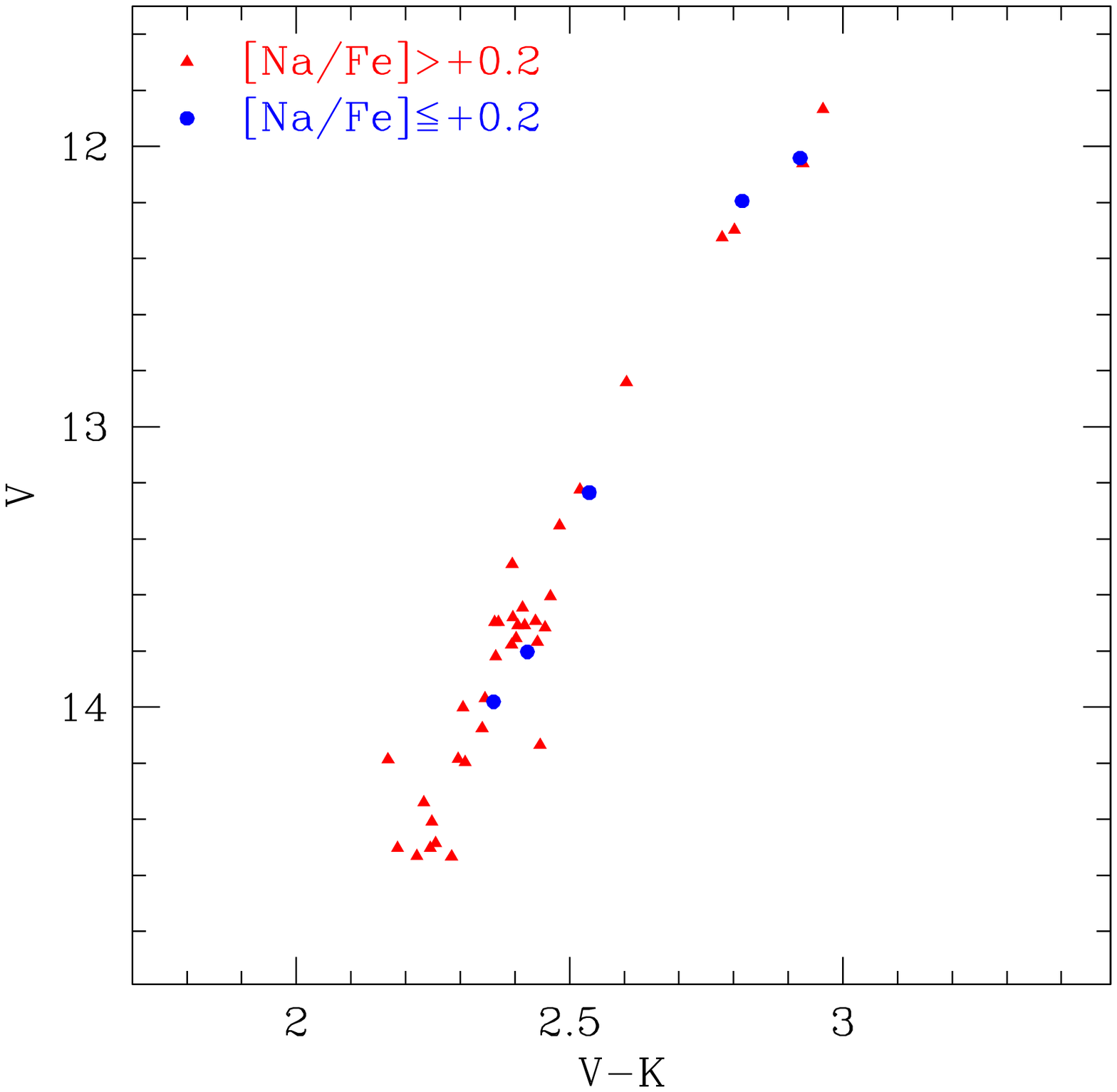}
\caption{Left panel: Str\"omgren $u$ vs $u-b$ CMD for the 42 stars in NGC~6752
in common with the unpublished photometry by Grundahl et al. (1999). Middle
panel: the same, but using the $V-K$ colour as abscissa. Right panel: $V$ vs
$V-K$ CMD. In all panels blue filled circles indicate first-generation stars (P
component) and red triangles second-generation stars (I component) in
NGC~6752.} 
\label{f:cmd3}
\end{figure*}

\begin{figure*}
\centering
\includegraphics[bb=45 290 580 570, clip, scale=0.69]{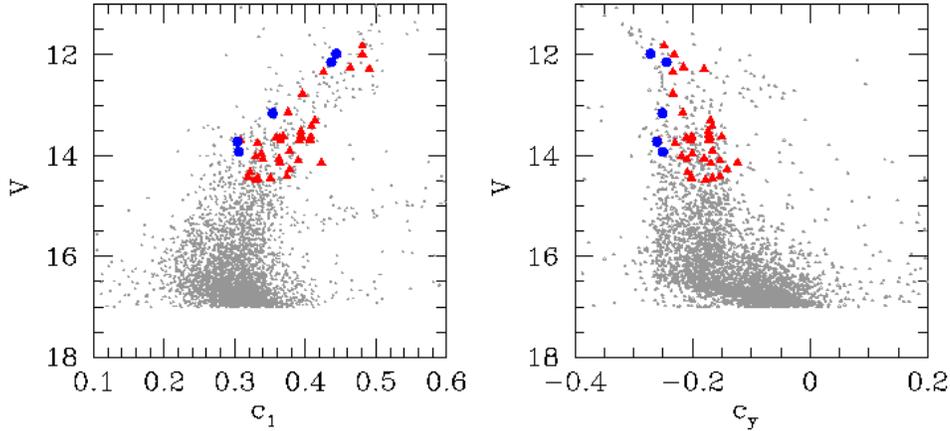}
\caption{Left and right panels: the $V$ vs $c_1$ and $V$ vs $c_y$ ,
respectively,  CMDs, as in Yong et al. (2008) using for NGC~6752 the unpublished
photometry by Grundahl et al. (1999). Filled (blue) circles and (red) triangles
are stars of the first and second-generations, respectively, as defined in this
work on the basis of their Na abundances alone.}
\label{f:c1cy}
\end{figure*}

Calibrated Johnson $U$ photometry is currently not available to us, but 
Str\"omgren photometry is for the programme cluster
NGC~6752. We cross-identified our sample in this cluster with
unpublished Str\"omgren photometry (Grundahl 1999, private
communication), finding 42 stars in common.  In the left panel of
Figure~\ref{f:cmd3}, these stars are plotted in the Str\"omgren $u, u-b$
CMD, with different symbols according to the division in stellar
populations in the previous section. We can see that the five
stars of the first stellar generation (P component) define a 
very tight sequence on the left ridge of the RGB,
while the other stars (all belonging to the I component) 
populate the remaining of the giant branch with a larger
dispersion. This is $not$ a temperature effect, due to systematic
differences in the effective temperature of first and second
generation stars, as can be inferred from the middle and right panels
in Figure~\ref{f:cmd3}. In the $u, V-K$ the separation of the two
sequence is less clear, and in the more classical $V, V-K$ CMD they
are virtually indiscernible.

Other evidence comes from the two Str\"omgren indexes $c_1$ and
$c_y$.  Yong et al. (2008) defined an empirical index $c_y$
designed to trace N abundances (its definition includes $c_1$, which in
turn uses the $u$ filter, where the effect of the NH band is
stronger), but removing temperature effects from the classical index
$c_1$. In Figure~\ref{f:c1cy} we superimpose the first and second-generation
stars we found in NGC~6752 (defined only from their abundances of Na)
on the diagrams from the whole Str\"omgren unpublished photometry for
this cluster, using both the $c_1$ index (left panel) and the newly
defined $c_y$ index, reproducing the same plots as in Yong et al. (2008,
their Figures 1 and 6).

Again, P stars define a very tight sequence, as expected for stars
born in a single burst of star formation in the still unpolluted
environment of the early GC. On the other hand, the
I component shows a much larger dispersion in both indexes,
as expected from stars born from matter resulting from a variable mix
of ejecta enriched in products of H-burning at high temperature and
pristine unpolluted gas.

\begin{figure}
\centering
\includegraphics[scale=0.45]{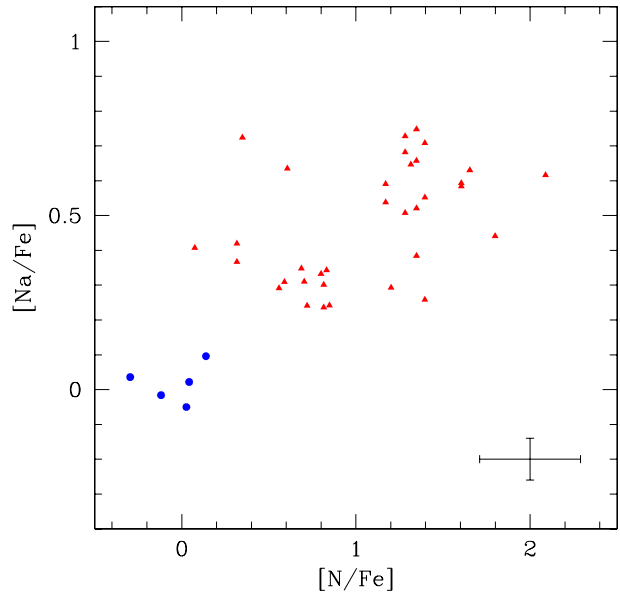}
\caption{Abundances of Na from our analysis (Paper II) vs [N/Fe] ratios derived
from the calibration by Yong et al. (2008) of the index $c_y$ for 42 stars in
NGC~6752 with Str\"omgren photometry. Errors bars are from Paper II (for Na) and
from the $rms$ scatter of the relation by Yong et al. (2008). Blue filled
circles are stars of the P component and red triangles are stars of the I
component.}
\label{f:nan}
\end{figure}

Finally, we can estimate the typical N content associated to the P and
I populations by using the empirical calibration given in Yong et al.
(2008) and derived exactly in NGC~6752. The results for our stars of
NGC~6752 are displayed in Figure~\ref{f:nan}, where the error bar in
[N/Fe] is the $rms$ scatter of the relation between [N/Fe] and $c_y$
quoted by Yong and collaborators (0.29 dex).  The average value of
[N/Fe] for the P component is about solar, [N/Fe]=-0.04 dex
($\sigma=0.17$ dex, 5 stars); for the I component, we derive a much
higher average value of [N/Fe]=+1.00 dex and a large scatter
($\sigma=0.50$ dex, 37 stars).

\begin{figure}
\centering
\includegraphics[bb=30 150 340 690, clip, scale=0.7]{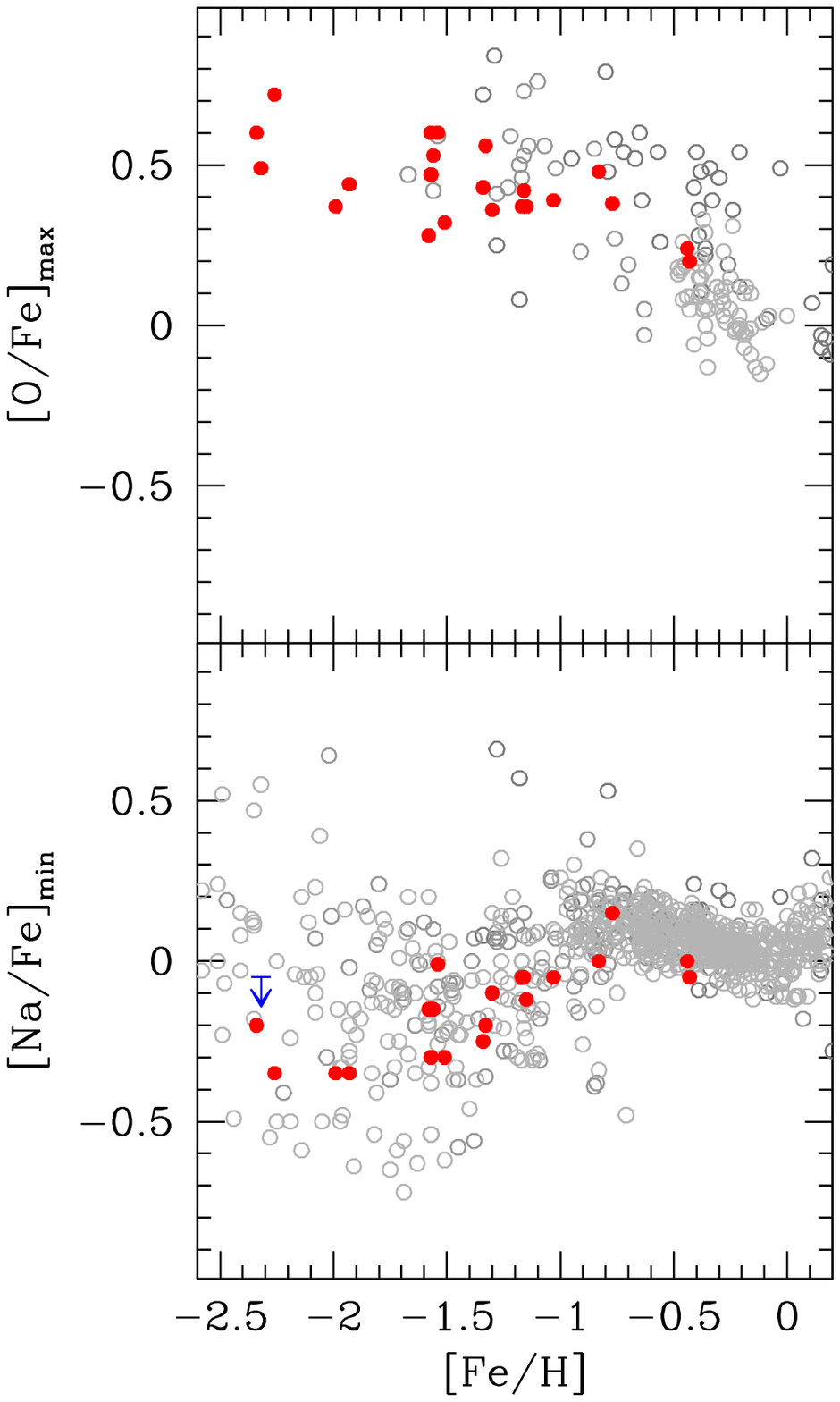}
\caption{Run of [O/Fe]$_{\rm max}$\ (upper panel) and [Na/Fe]$_{\rm min}$\ 
(lower panel) with [Fe/H] for the GCs of our sample. These should 
represent the run of original O and Na abundances for these clusters. 
The grey, open symbols, represent field stars taken from Fulbright, McWilliam \& Rich (2007), Venn et al. (2004),  Gratton et al. (2003),  Reddy et al. (2003).}
\label{f:omaxnamin}
\end{figure}

\section{A dilution model for the Na-O anticorrelation and the shape
of the Na-O anticorrelation}

We do not have a satisfactory model yet for the mechanism responsible of
the Na-O anticorrelation, and even the astrophysical site is currently
debated (fast-rotating massive stars vs massive AGB stars undergoing
hot bottom burning: see Decressin et al. 2007; Ventura et al. 2001). A
simple approach is to assume (i) that within each cluster there is a
unique mechanism that produces some given amount of sodium and
destroys almost all O (transforming it into N); and (ii) that the
processed material is then mixed with a variable amount of pristine
material. A similar dilution model has been successfully used to
explain many features of the Na-O anticorrelation (see discussion in
Prantzos et al. 2007).  Once the compositions of the pristine and
processed material are set (e.g., by the extremes of the observed
distributions), an appropriate dilution factor may be determined for
each star (either from O or Na abundances).

In this model the logarithmic abundance of an element [X] for a given
dilution factor $dil$ is given by:
\begin{equation} 
[X] = \log{[(1-dil)~10^{\rm [Xo]} + dil~10^{\rm[Xp]}]},
\end{equation}
where [Xo] and [Xp] are the logarithmic abundance of the element in
the original and processed material. In principle, [Xo] and [Xp] could
be derived from observations for both O and Na. We may adopt for [Xo]
the maximum observed abundance of O ([O/Fe]$_{\rm max}$) and the
minimum observed abundance of Na ([Na/Fe]$_{\rm min}$), and for [Xp]
the minimum observed abundance of O ([O/Fe]$_{\rm min}$) and the
maximum observed abundance of Na ([Na/Fe]$_{\rm max}$). Practically,
we derived minimum O and Na abundances by visual inspection of the
observed distributions, while we obtained the maximum Na and O
abundances by minimising the r.m.s. of points due to individual stars
along dilution fitting relations \footnote{When doing this exercise,
  we considered upper limits as actual detections. Also, in the case
  of NGC~6441 we neglected the two stars with the largest Na
  abundances, which clearly stand out with respect to the relation given by the
  other stars.}.  Table~\ref{t:dilution} gives the minimum and maximum
O and Na abundances we obtained for the 19 clusters in our programme.
In many cases we can only derive upper limits to
[O/Fe]$_{\rm min}$.  This is surely the case for the most metal-poor
GCs ([Fe/H]$<-1.7$), where we may grossly overestimate the
[O/Fe]$_{\rm min}$. We explicitly indicate this in the second
column of Table~\ref{t:dilution}. Also, [Na/Fe]$_{\rm min}$ can be
overestimated for the most metal-poor GCs, which we think this may be the case
for M15 (NGC~7078).

\begin{table}
\centering
\caption[]{Minimum and maximum abundances of O and Na from our 
dilution model}
\scriptsize
\begin{tabular}{rcccccc}
\hline
 NGC & [O/Fe]$_{\rm min}$ &[Na/Fe]$_{\rm min}$ &  n  &[O/Fe]$_{\rm max}$&[Na/Fe]$_{\rm max}$&rms \\
\hline
 104 &  -0.4 & 0.15 & 114 & $0.38\pm 0.08$ & $0.74\pm 0.06$ & 0.08 \\
 288 &  -0.5 &-0.10 &  70 & $0.36\pm 0.18$ & $0.71\pm 0.18$ & 0.18 \\ 
1904 &  -0.6 &-0.15 &  48 & $0.28\pm 0.09$ & $0.72\pm 0.07$ & 0.10 \\ 
2808 &  -1.0 &-0.12 &  98 & $0.37\pm 0.07$ & $0.56\pm 0.04$ & 0.10 \\ 
3201 &  -0.8 &-0.30 & 100 & $0.32\pm 0.10$ & $0.60\pm 0.09$ & 0.15 \\ 
4590 & $<0.0$&-0.35 &  48 & $0.72\pm 0.20$ & $0.53\pm 0.13$ & 0.13 \\ 
5904 &  -0.7 &-0.25 & 124 & $0.43\pm 0.13$ & $0.60\pm 0.10$ & 0.14 \\ 
6121 &  -0.2 &-0.05 &  88 & $0.37\pm 0.07$ & $0.74\pm 0.08$ & 0.07 \\ 
6171 &  -0.3 &-0.05 &  30 & $0.39\pm 0.09$ & $0.69\pm 0.07$ & 0.08 \\ 
6218 &  -0.4 &-0.20 &  74 & $0.56\pm 0.11$ & $0.67\pm 0.07$ & 0.13 \\ 
6254 &  -0.4 &-0.30 &  87 & $0.47\pm 0.09$ & $0.56\pm 0.10$ & 0.12 \\ 
6388 &  -0.6 & 0.00 &  32 & $0.24\pm 0.11$ & $0.67\pm 0.05$ & 0.10 \\ 
6397 & $<0.0$&-0.35 &  16 & $0.37\pm 0.09$ & $0.71\pm 0.23$ & 0.06 \\ 
6441 &  -0.4 &-0.05 &  27 & $0.20\pm 0.11$ & $0.80\pm 0.10$ & 0.12 \\ 
6752 & $-0.4$&-0.15 &  98 & $0.53\pm 0.13$ & $0.65\pm 0.07$ & 0.14 \\ 
6809 &$<-0.2$&-0.35 &  84 & $0.44\pm 0.14$ & $0.69\pm 0.09$ & 0.12 \\ 
6838 &   0.0 & 0.00 &  42 & $0.48\pm 0.10$ & $0.76\pm 0.16$ & 0.09 \\ 
7078 &$<-0.1$&$<-0.05$&  33 & $0.49\pm 0.09$ & $0.70\pm 0.09$ & 0.09 \\ 
7099 &$<-0.2$&-0.20 &  29 & $0.60\pm 0.15$ & $0.76\pm 0.14$ & 0.14 \\ 
\hline
\end{tabular}
\label{t:dilution}
\end{table}

As mentioned above, the minimum Na and maximum O abundances in each
cluster represent the original Na and O composition of the cluster. It
is interesting to plot their runs with [Fe/H] and to compare them with
the runs observed in field halo stars (see Fig.~\ref{f:omaxnamin}). The
upper panel of this figure indicates that GCs generally started from
high values of the O abundances of [O/Fe]$\sim 0.35\div 0.5$,
implying a marginal contribution (if any) by type Ia SNe to their
original composition. Only the two most metal-rich clusters (NGC~6388
and NGC~6441), both at ${\rm [Fe/H]}\sim -0.4$, have a moderate excess
of O ([O/Fe]$\sim 0.2$): these clusters are well beyond the knee of the
[O/Fe] run observed for field stars. Also the run for [Na/Fe]$_{\rm
  min}$ with [Fe/H] closely reflects that observed among metal-poor
stars. We conclude that the original composition of GCs reflected the
typical composition of the field halo material.

\begin{figure*}
\centering
\includegraphics[bb=20 390 570 690,clip,scale=0.7]{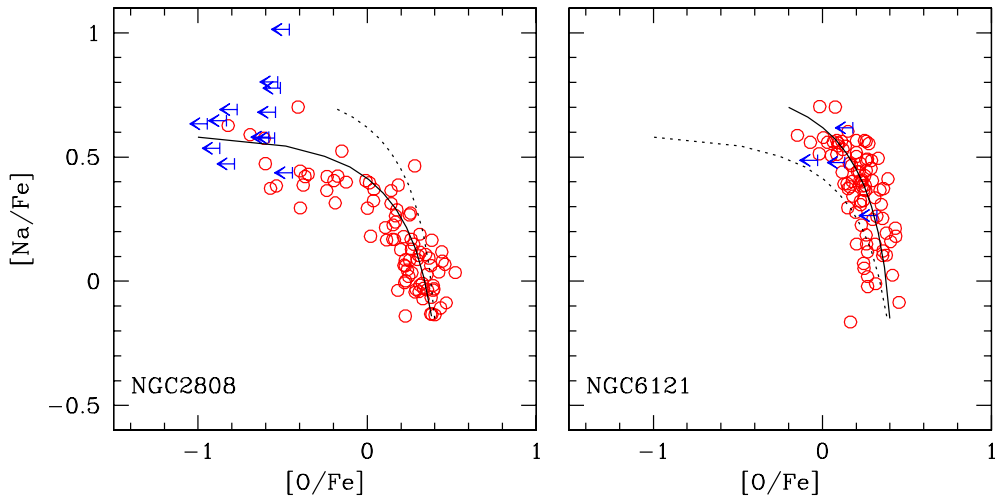}  
\caption{The O/Na anticorrelation in NGC~2808 (left panel) and M~4 (NGC~6121, right
panel). Red circles represent stars with actual measures of the O abundances,
while blue arrows represent those stars for which only upper limits were
obtained. Overlying lines are the results of our dilution model for the two
clusters, respectively. }
\label{f:fignaodil}
\end{figure*}

\begin{figure}
\centering
\includegraphics[bb=20 380 360 686,clip, scale=0.7]{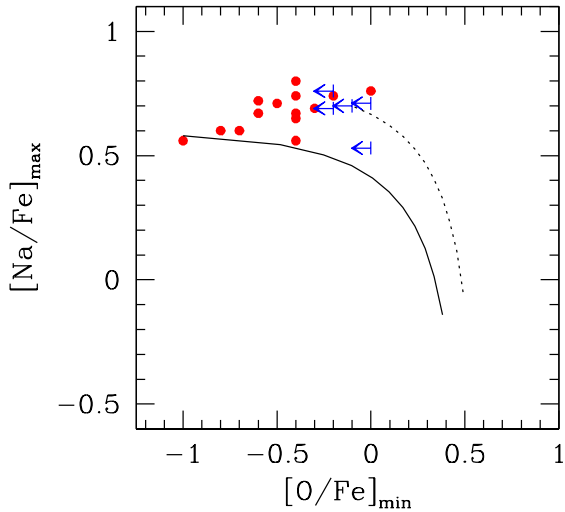} 
\caption{Run of [Na/Fe]$_{\rm max}$\ with [O/Fe]$_{\rm min}$\ for the GCs of 
our sample.
Arrows represent upper limits and the two lines are the dilution relations
for NGC~2808 and NGC~6121 shown in the previous figure.}
\label{f:ominnamax}
\end{figure}

\begin{figure}
\centering
\includegraphics[bb=20 380 360 686, clip, scale=0.7]{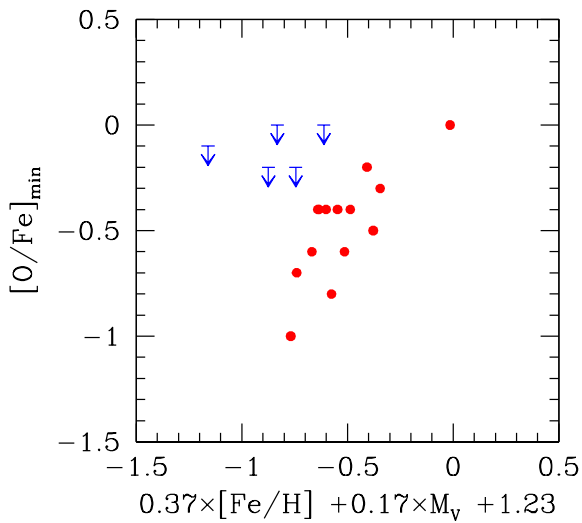} 
\caption{Run of [O/Fe]$_{\rm min}$\ for the GCs of our sample as a function of
a linear combination of metallicity [Fe/H] and cluster luminosity $M_V$. This
relation was computed by excluding the five most metal-poor GCs, indicated by
arrows.}
\label{f:ominrelaz}
\end{figure}

The minimum O and maximum Na abundances in each cluster determine the
slope of the O/Na anticorrelation. If the polluters were massive AGB
stars, these two quantities would be expected to be anticorrelated,
depending on the average mass and metallicity of the polluters; hence,
[O/Fe]$_{\rm min}$ and [Na/Fe]$_{\rm max}$ might change from cluster
to cluster.  This is indeed the case: for instance, NGC~2808, with
[Na/Fe]$_{\rm max}=0.58\pm 0.03$\, has a O/Na anticorrelation 
clearly
flatter than M 4 ([Na/Fe]$_{\rm max}=0.70\pm 0.11$; see
Figure~\ref{f:fignaodil}), in spite of the fact that these two
clusters have very similar values of [Fe/H]. This suggests that the
average mass of the polluters may be larger in NGC~2808 than in M 4.
Searching for general trends, we plotted 
the run of [Na/Fe]$_{\rm max}$ with [O/Fe]$_{\rm min}$ 
in Figure~\ref{f:ominnamax}. If we neglect
the upper limits (which do not provide useful information here), we
find that these two quantities are correlated, as expected for
massive AGB polluters. However, the observed slope is quite different
from model expectations, the variation in [O/Fe]$_{\rm min}$ being
much greater than expected. This might indicate some flaws in the model
(e.g. in the treatment of convection and/or on the adopted value for
the relevant nuclear reaction cross sections). However (still
excluding the most metal-poor clusters, which only provide not very
constraining upper limits), we find that [O/Fe]$_{\rm min}$ is closely
correlated with a linear combination of metallicity [Fe/H] and cluster
luminosity $M_V$, the mean relation being
\begin{equation} 
{\rm [O/Fe]}_{\rm min} = (0.366\pm 0.134) {\rm [Fe/H]} + (0.168\pm
0.044) M_V + (1.23\pm 0.17), 
\end{equation} 
with a linear correlation correlation coefficient of $r=0.77$ (over 14
GCs), which is highly significant (see also Figure~\ref{f:ominrelaz}).
The correlation with the cluster absolute magnitude suggests that the
average mass of the polluting stars is correlated with the cluster's
absolute magnitude (or, more likely, with the mass of the cluster at the
epoch of formation, of which its current $M_V$\ value can be a proxy).
It is unfortunate that current AGB models are not yet able to provide a
good calibration of the polluting mass, because this would provide us
with the typical timescale for the formation of the second-generation,
a crucial piece of information in modelling early phases of cluster evolution.

\section{Summary}

In this paper we have derived atmospheric parameters and elemental
abundances of iron, oxygen, and sodium for 1582 {\it bona fide} member
red giant stars in 15 Galactic GCs with different global
parameters (metallicity, masses, HB morphology, etc.). We derived our
abundances from $EW$s measured on high-resolution
FLAMES/GIRAFFE spectra; the $EW$s are corrected to a system defined by
higher resolution FLAMES/UVES spectra (presented in the companion
Paper VIII).  We added stars analysed in the previous studies within the 
project to the sample of the present paper, and the resulting sample
was completed with the UVES dataset. We have a grand total of 1235
stars with homogeneous Na and O abundances in 19 clusters, the largest
sample of its kind ever collected. This huge database allows 
tracing the Na-O anticorrelation for each GC, the typical signature
of operation of proton-capture chains in high-temperature H-burning in
an early generation of -now extinct- massive stars. This classical
sign of large star-to-star abundance variations {\em is present
  in all clusters studied to date, so it must be fundamentally
  related to the mechanisms of formation and early evolution of GCs}.
For some of the clusters in our sample the Na-O anticorrelation is
detected here for the first time.

Our homogeneous abundances are used to provide a chemically tag of
multiple stellar populations and allow us to separate and $quantify$
the fraction of first and second-generation stars in globular
clusters. A component P is identified with stars populating
(in the Na-O plane) the locus occupied by field stars of similar
metallicity, showing only the chemical pattern from supernovae
nucleosynthesis. This P component is present in all clusters, at 
a level averaging from about 30 up to (in a few cases) 50\% : no cluster
is found completely lacking the pristine stellar component. 
This is at variance with the suggestion (D'Antona and Caloi 2008)
that some clusters (e.g.  NGC~6397) are only composed of second-generation stars.

The remaining stars are second-generation stars, formed by the gas
pool polluted by intermediate and/or massive first-generation stars.
According to the degree of changes in O and Na, we could separate this
second-generation into an I and E populations.
The I component represents the bulk of the clusters' present
population, including up to 60-70\% of currently observed cluster
stars. The E population is not present in all clusters and is
more easily found in very massive clusters. However, this is a
necessary but not sufficient condition: massive clusters such as
47~Tuc (NGC~104) and maybe M~15 (NGC~7078) do not harbour a significant fraction of stars
with heavily modified chemical composition.

We found a tendency for I stars to be more concentrated
toward the cluster centre than P stars, but the significance
of this finding might be somewhat biased by our
likely observing dynamically different regions in different clusters
(due to the combination of cluster parameters and size on the sky and
to the mechanical limitations of the fibre positioner of FLAMES).
Although there are hints of a different spatial distribution of the
three P,I,E cluster populations, further observations and larger samples
of stars are needed, especially in the smaller clusters.

Using Str\"omgren photometry we verified in NGC~6752 that stars of the
first-generation are also N-poor, while stars of the second-generation
(the intermediate  component) are N-rich. The N content affects blue
colours (such as the $u-b$) through the $u$ band flux; this causes the P 
stars to lie along a tight
sequence on the blue of the RGB, while the I stars, composed of a mix
of polluted, N-enriched matter, and of pristine gas, populate a wider
part of the CMD.

Finally, the comparison of the observed Na-O anticorrelation with
dilution sequences has allowed us to (i) determine the original O and
Na abundances; (ii) show that these anticorrelations differ
systematically from cluster to cluster, the maximum Na and minimum O
abundances being correlated, in qualitative but not quantitative
agreement with nucleosynthesis prediction for massive AGB stars; and
(iii) find that the slope of the Na-O anticorrelation is driven by
metallicity and cluster absolute magnitude (or mass). When compared
with the nucleosynthesis predictions, this suggests that the average
mass of polluting stars is anticorrelated with total cluster mass.

\begin{acknowledgements}
  This publication makes use of data products from the Two Micron All
  Sky Survey, which is a joint project of the University of
  Massachusetts and the Infrared Processing and Analysis
  Center/California Institute of Technology, funded by the National
  Aeronautics and Space Administration and the National Science
  Foundation. This work was partially funded by the Italian MIUR under
  PRIN 2003029437. We also acknowledge partial support from the grant
  INAF 2005 ``Experimenting nucleosynthesis in clean environments''.
  We warmly thank the referee, Mike Bessell, for the careful reading of this
  manuscript and for his suggestions improving the paper.
  S.L. is grateful to the DFG cluster of excellence ``Origin and
  Structure of the Universe'' for partial support.  M.B. acknowledges the
  financial support of INAF through the PRIN-2007 grant CRA
  1.06.10.04. We also wish to thank F. De Angeli for preparatory work
  and C. Corsi, L. Pulone, and F.
  Grundahl for their unpublished photometry. This research
  has made use of the SIMBAD database, operated at CDS, Strasbourg,
  France and of NASA's Astrophysical Data System.
\end{acknowledgements}

\begin{appendix}
\section{Error estimates}
\subsection{Individual (star-to-star) errors} In the following discussion we
focus our attention on  individual (i.e. star-to-star) errors in the
derived abundances that are relevant when discussing the internal spread of
abundance within a cluster, which is our main aim.  As shown in previous works
(see e.g. Paper IV, Paper V), the main error sources are  those in temperature,
microturbulent velocity and $EW$s.  The effects of
errors in surface gravities and in the adopted model metallicity are negligible
in the total error budget.

The error estimate can be split into 3 steps.

\paragraph{Sensitivities of abundance ratios to atmospheric parameters.}  The
first step in our error analysis is to evaluate the sensitivity of the derived
abundances to the adopted atmospheric parameters. These sensitivities were
obtained by repeating our abundance analysis by changing only one atmospheric
parameter each time.

Notice that {\it at least} two typical cases (a cool and a warm star) are
required, because cool and warm stars are in two different regimes, in warm
stars Fe is mainly ionised, while in cool stars Fe is mainly neutral. For this
reason, sensitivities to variations in effective temperatures and surface
gravities are different in the two cases.  In our case, this exercise was done
on $all$ the stars in each cluster. Afterward, we adopt the  sensitivity in each
parameter as the one corresponding to the average of all the sample (separately
for each cluster).

The amount of the variations in the atmospheric parameters and the resulting 
response in abundance changes of Fe, O, and Na (the sensitivities) are shown in
Table~\ref{t:sensitivity}.

\begin{table*}
\small
\caption{Sensitivities of abundance ratios to errors in the atmospheric
parameters}
\begin{tabular}{lllllcrrrr}
\hline
        & \multicolumn{4}{c}{$\Delta T_{\rm eff}=50$~K} &\multicolumn{1}{c}{} &\multicolumn{4}{c}{$\Delta V_t = +0.1$~km/s} \\
	\cline{2-5} \cline{7-10}\\

cluster &$\Delta$[Fe/H]I&$\Delta$[Fe/H]II&$\Delta$[O/Fe]&$\Delta$[Na/Fe]&    &$\Delta$[Fe/H]I&$\Delta$[Fe/H]II&$\Delta$[O/Fe]&$\Delta$[Na/Fe] \\

NGC 104 & +0.033& $-$0.049& $-$0.025&   +0.009&   &$-$0.034& $-$0.016& +0.035& +0.008 \\
NGC 288 & +0.055& $-$0.026& $-$0.044& $-$0.016&   &$-$0.027& $-$0.012& +0.029& +0.018 \\
NGC~1904& +0.060& $-$0.022& $-$0.042& $-$0.022&   &$-$0.027& $-$0.008& +0.029& +0.021 \\
NGC~3201& +0.059& $-$0.022& $-$0.044& $-$0.021&   &$-$0.022& $-$0.009& +0.024& +0.020 \\
NGC~4590& +0.049& $-$0.007& $-$0.023& $-$0.023&   &$-$0.011& $-$0.005& +0.014& +0.009 \\
NGC~5904& +0.057& $-$0.029& $-$0.043& $-$0.016&   &$-$0.028& $-$0.011& +0.028& +0.019 \\
NGC~6121& +0.049& $-$0.034& $-$0.038& $-$0.006&   &$-$0.031& $-$0.011& +0.033& +0.018 \\
NGC~6171& +0.045& $-$0.037& $-$0.035& $-$0.002&   &$-$0.026& $-$0.013& +0.028& +0.015 \\
NGC~6254& +0.055& $-$0.014& $-$0.037& $-$0.021&   &$-$0.018& $-$0.006& +0.020& +0.015 \\
NGC~6388& +0.015& $-$0.073& $-$0.006&   +0.023&   &$-$0.044& $-$0.023& +0.045& +0.014 \\
NGC~6397& +0.047& $-$0.007& $-$0.034& $-$0.021&   &$-$0.009& $-$0.003& +0.010& +0.009 \\
NGC~6808& +0.062& $-$0.016& $-$0.042& $-$0.028&   &$-$0.015& $-$0.004& +0.017& +0.012 \\
NGC~6838& +0.052& $-$0.038& $-$0.046& $-$0.011&   &$-$0.032& $-$0.016& +0.033& +0.013 \\
NGC~7078& +0.050& $-$0.011& $-$0.026& $-$0.023&   &$-$0.008& $-$0.002& +0.011& +0.007 \\
NGC~7099& +0.047& $-$0.010& $-$0.021& $-$0.021&   &$-$0.008& $-$0.002& +0.010& +0.006 \\

\\
\hline
        & \multicolumn{4}{c}{$\Delta \log g = +0.2$~dex} &\multicolumn{1}{c}{} &\multicolumn{4}{c}{$\Delta$ [A/H] = +0.1~dex} \\
	\cline{2-5} \cline{7-10}\\

cluster &$\Delta$[Fe/H]I&$\Delta$[Fe/H]II&$\Delta$[O/Fe]&$\Delta$[Na/Fe]&    &$\Delta$[Fe/H]I&$\Delta$[Fe/H]II&$\Delta$[O/Fe]&$\Delta$[Na/Fe] \\

NGC 104 &   +0.015& +0.107& +0.070& $-$0.054&	&   +0.009& +0.037& +0.027& $-$0.001 \\
NGC 288 & $-$0.009& +0.087& +0.090& $-$0.025&	& $-$0.007& +0.020& +0.034& $-$0.004 \\
NGC~1904& $-$0.011& +0.083& +0.090& $-$0.025&	& $-$0.010& +0.018& +0.036& $-$0.005 \\
NGC~3201& $-$0.009& +0.084& +0.089& $-$0.019&	& $-$0.009& +0.021& +0.038& $-$0.001 \\
NGC~4590& $-$0.005& +0.073& +0.079& $-$0.015&	& $-$0.003& +0.005& +0.016&   +0.003 \\
NGC~5904& $-$0.005& +0.087& +0.086& $-$0.029&	& $-$0.007& +0.022& +0.037& $-$0.007 \\
NGC~6121& $-$0.004& +0.093& +0.089& $-$0.033&	& $-$0.002& +0.025& +0.034& $-$0.011 \\
NGC~6171& $-$0.003& +0.096& +0.090& $-$0.038&	&   +0.003& +0.028& +0.032& $-$0.003 \\
NGC~6254& $-$0.009& +0.079& +0.087& $-$0.020&	& $-$0.009& +0.013& +0.030& $-$0.000 \\
NGC~6388&   +0.032& +0.117& +0.052& $-$0.071&	&   +0.018& +0.042& +0.018& $-$0.001 \\
NGC~6397& $-$0.003& +0.073& +0.081& $-$0.012&	& $-$0.002& +0.007& +0.026& $-$0.001 \\
NGC~6808& $-$0.009& +0.078& +0.084& $-$0.027&	& $-$0.011& +0.014& +0.034& $-$0.002 \\
NGC~6838&   +0.005& +0.099& +0.081& $-$0.042&	&   +0.007& +0.032& +0.029& $-$0.002 \\
NGC~7078& $-$0.008& +0.072& +0.078& $-$0.017&	& $-$0.002& +0.006& +0.018&   +0.005 \\
NGC~7099& $-$0.004& +0.073& +0.076& $-$0.014&	& $-$0.002& +0.006& +0.015&   +0.003 \\

\\
\hline
\end{tabular}
\label{t:sensitivity}
\end{table*}

\paragraph{Errors in atmospheric parameters} The next step is to evaluate the
actual errors in the atmospheric parameters. The individual star errors are
those that show up when we compare abundances obtained from different stars in
the same cluster. A detailed and more wordy discussion of how they can be
estimated is given in Paper IV; here we only provide a schematic description.
Results are given in columns labelled 2 to 6 of Table~\ref{t:errpar}.

\begin{itemize}
\item internal error in $T_{\rm eff}$ are estimated from the slope of the
relation between $T_{\rm eff}(V-K)$ from the Alonso et al. calibration and the
$V$ or $K$ magnitude, assuming an error of 0.02 mag.
\item the error in the micro-turbulent velocity is estimated by the change in
$v_t$ required to vary the slope of the expected line strength vs abundances
relation by 1$\sigma$; this value was derived as the quadratic mean of the
1 $\sigma$ errors in the slope of the relation between abundance and expected
line strength for all stars with enough Fe~I lines measured.
\item to estimate errors in the measurement of $EW$s we selected 
 a subset of stars with  more than 15 measured Fe lines in each
cluster (this number
dropped down to 10 or 6 for the most metal-poor clusters). The average $rms$
scatter in Fe abundance for these stars, divided by the square root of the
typical average  number of measured lines, provides the typical internal errors
listed in Table~\ref{t:errpar}, column label (5).
\end{itemize}

\paragraph{Estimate of error in abundances} Once estimates of the individual
star errors in the atmospheric parameters are available (Table~\ref{t:errpar}),
they may be multiplied for the sensitivities  of abundances to variations in the
individual parameters (Table~\ref{t:sensitivity}) to derive  their contribution
to the total individual star errors, listed in Tab~\ref{t:errabu}. 

Total errors, computed by summing in quadrature only the dominant terms  (due to
$T_{\rm eff}$, $v_t$ and $EW$s), or including all the contributions, are
reported in  Table~\ref{t:errabu}, in Cols. 8 and 9 respectively, for iron and
for the other two elements O and Na. From this table one can also appreciate how
negligible is to include of error sources due to gravity and model metal
abundance.

In almost all clusters the observed scatter (col. 5 in Table~\ref{t:quantefe}) 
is formally lower than the total star-to-star error, which might indicate 
that the errors are slightly overestimated.

In summary, our abundance analysis and error estimate allow us to conclude that
each of the clusters of the present project shows a high degree of homogeneity as
far as the global metallicity is concerned, since we do not find any
statistically significant intrinsic spread in [Fe/H].

\begin{table*}
\small
\caption{Star-to-star (individual) errors and cluster errors in atmospheric
parameters and in the EWs}
\begin{tabular}{lrllllcrrrr}
\hline
     & \multicolumn{5}{c}{star-to-star errors}  &\multicolumn{1}{c}{} &\multicolumn{4}{c}{cluster errors} \\
	\cline{2-6} \cline{8-11}\\

cluster &$T_{\rm eff}$& $\log g$ & [A/H] & $v_t$ & EW    &   & $T_{\rm eff}$& $\log g$ & [A/H] & $v_t$\\
        &    (K)      &  dex     & (dex) & (km/s)&(dex)  &   &  (K)         &  dex     & (dex) & (km/s) \\
	&    (1)      &  (2)     & (3)   & (4)   &(5)    &   &  (6)         &  (7)     & (8)   & (9)   \\
	\cline{2-6} \cline{8-11}\\ 
NGC 104 &  	6     &  0.042   & 0.032 &  0.11 & 0.025 &   &    40        &  0.059   & 0.026 & 0.009 \\
NGC 288 &  	6     &  0.041   & 0.042 &  0.28 & 0.025 &   &    63        &  0.061   & 0.070 & 0.027 \\
NGC~1904&  	5     &  0.041   & 0.036 &  0.20 & 0.027 &   &    57        &  0.060   & 0.069 & 0.026 \\
NGC~3201&  	4     &  0.041   & 0.049 &  0.19 & 0.028 &   &    62        &  0.061   & 0.073 & 0.016 \\
NGC~4590&  	4     &  0.041   & 0.071 &  0.28 & 0.037 &   &    69        &  0.061   & 0.068 & 0.025 \\
NGC~5904&      12     &  0.041   & 0.023 &  0.11 & 0.024 &   &    54        &  0.060   & 0.062 & 0.009 \\
NGC~6121&  	4     &  0.041   & 0.025 &  0.12 & 0.022 &   &    54        &  0.060   & 0.053 & 0.012 \\
NGC~6171&  	2     &  0.041   & 0.044 &  0.21 & 0.025 &   &    26        &  0.057   & 0.026 & 0.037 \\
NGC~6254&  	4     &  0.041   & 0.053 &  0.13 & 0.026 &   &    67        &  0.061   & 0.074 & 0.011 \\
NGC~6388&  	9     &  0.043   & 0.078 &  0.19 & 0.037 &   &    57        &  0.061   & 0.028 & 0.032 \\
NGC~6397&  	4     &  0.041   & 0.039 &  0.34 & 0.038 &   &    64        &  0.060   & 0.060 & 0.028 \\
NGC~6808&  	5     &  0.041   & 0.044 &  0.20 & 0.027 &   &    58        &  0.060   & 0.072 & 0.016 \\
NGC~6838&  	5     &  0.041   & 0.034 &  0.10 & 0.023 &   &    45        &  0.059   & 0.048 & 0.016 \\
NGC~7078&  	5     &  0.041   & 0.061 &  0.33 & 0.030 &   &    67        &  0.061   & 0.067 & 0.036 \\
NGC~7099&  	5     &  0.041   & 0.046 &  0.41 & 0.034 &   &    71        &  0.061   & 0.067 & 0.051 \\

\\
\hline
\end{tabular}
\begin{list}{}{}
\item[(1)] slope relation $T_{\rm eff}$(V-K)$_{Alonso}$ $vs$ mag V or K + 0.02 mag error in V or K
\item[(2)] slope $\log g$ $vs$ mag V or K +0.02 mag error + 10\% variation in mass
\item[(3)] $rms$ scatter in [Fe/H] of all analysed stars
\item[(4)] quadratic mean of 1$\sigma$ errors in the slope abundances Fe~I/line strength (minus systematic components) 
from stars with a large enough number of Fe~I lines
\item[(5)] ($rms$ in Fe~I for stars with enough lines) divided the square root of typical number of lines
\item[(6)] slope relation $T_{\rm eff}$(V-K)$_{Alonso}$ $vs$ (V-K)$_0$ +0.02 error in E(B-V) + 0.02 mag error
in V-K colours zero point.
\item[(7)] 0.1 mag error in modulus + systematic error in $T_{\rm eff}$ +10\% error in mass
\item[(8)] statistical error+systematic error in $T_{\rm eff}$+systematic error in $\log g$+systematic error in $v_t$
\item[(9)] internal error in $v_t$ divided the square root of N$_{stars}$
\end{list}
\label{t:errpar}
\end{table*}

\subsection{Cluster (systematic) errors}

In this section we examine the errors that are systematic for all stars in
a cluster, but are different for the various clusters considered in this series
of paper on the Na-O anticorrelation and HB. Hence, they will have no effect on
the star-to-star scatter, but will produce scatter in the relations involving
different clusters. We proceed following the same order as considered for the
individual star errors.
\\

(i) {\bf $T_{\rm eff}$}. As mentioned before, effective temperatures were
derived from magnitudes, adopting a mean relation between $V$\ or $K$\
magnitudes and $T_{\rm eff}(V-K)$, which in turn are derived from  $V-K$ colours
using the calibration by Alonso et al. (1999). The $V-K$\ colours to be used
here are of course the dereddened colours (in the TCS system). Errors in the
assumed reddening (and on the zero point of the photometric scales) will cause a
systematic shift in the $T_{\rm eff}$'s. 

The reddening estimate we used are from Harris (1996). Assuming an uncertainties
of 0.02 mag in $E(B-V)$ this implies an uncertainty in $E(V-K)$\ of 0.02$\times
2.75=0.055$ mag. We may then estimate the cluster uncertainty in the  $T_{\rm
eff}$'s by multiplying the uncertainty in $E(V-K)$\ for the slope of the
relation between  $T_{\rm eff}$ and $V-K$\ derived in each cluster.

Including a (conservative) estimate of 0.02 mag error in the zero point of the
$V-K$\ colours, and summing this error quadratically to the error in the
reddening, the errors come out to be as in column labelled (6) in
Table~\ref{t:errpar}. \\

(ii) {\bf $\log{g}$}. Errors in surface gravity might be obtained by propagating
uncertainties in distance modulus (about 0.1 mag), stellar mass (a conservative
10\%) and the above systematic errors in effective temperature. The quadratic
sum results  in errors listed in column (7) of in Table~\ref{t:errpar}, and is very
similar for all clusters. \\

(iii) {\bf $v_t$} The systematic error in $v_t$ is simply the internal error in
$v_t$ divided for the square root of the number of stars (in each cluster). \\

(iv)  [A/H]. The cluster error we consider here is given by the
quadratic sum of four terms: the first 3 are the systematic contribution
estimated above multiplied for the appropriate sensitivities in
Table~\ref{t:sensitivity}. The last one is simply the statistical errors of 
individual abundance determinations ($rms$ scatter divided the square root of
the number of stars used in each cluster).  \\

Total systematic errors related to individual clusters are listed in 
Table~\ref{t:errpar}.

\begin{table*}
\scriptsize
\caption{Error in element ratios due to star-to-star errors in atmospheric
parameters and in the EWs}
\begin{tabular}{lllllrlcccl}
\hline
& \multicolumn{6}{c}{errors in abundances due to:} &\multicolumn{1}{c}{} &\multicolumn{2}{c}{total star-to-star error} &\\
  \cline{2-7} \cline{9-10}\\
  
            & $T_{\rm eff}$& $\log g$ & [A/H]   & $v_t$  &$<$nr$>$ & EW  &  & $T_{\rm eff}$+$v_t$+EW &     all &\\
$[$Fe/H$]$I &     +0.004   &  +0.001  &  +0.003 &$-$0.039& 30   & 0.025  &  &  	 0.046  	     &   0.047 &NGC 104\\
$[$Fe/H$]$II&   $-$0.006   &  +0.004  &  +0.011 &$-$0.017&  3   & 0.078  &  &  	 0.080  	     &   0.081 &\\
$[$O/Fe$]$  &   $-$0.003   &  +0.003  &  +0.009 &  +0.039&  1   & 0.135  &  &  	 0.141  	     &   0.141 &\\
$[$Na/Fe$]$ &     +0.001   &$-$0.002  &  +0.000 &  +0.010&  3   & 0.078  &  &  	 0.079  	     &   0.079 &\\
$[$Fe/H$]$I &     +0.007   &$-$0.002  &$-$0.002 &$-$0.076& 31   & 0.025  &  &  	 0.080  	     &   0.080 &NGC 288\\
$[$Fe/H$]$II&   $-$0.003   &  +0.018  &  +0.018 &$-$0.034&  2   & 0.098  &  &  	 0.104  	     &   0.106 &\\
$[$O/Fe$]$  &   $-$0.005   &  +0.018  &  +0.018 &  +0.081&  1   & 0.138  &  &  	 0.160  	     &   0.162 &\\
$[$Na/Fe$]$ &   $-$0.002   &$-$0.005  &$-$0.005 &  +0.050&  2   & 0.098  &  &  	 0.110  	     &   0.110 &\\
$[$Fe/H$]$I &     +0.006   &$-$0.002  &$-$0.004 &$-$0.054& 25   & 0.027  &  &  	 0.061  	     &   0.061 &NGC 1904\\
$[$Fe/H$]$II&   $-$0.002   &  +0.017  &  +0.006 &$-$0.016&  2   & 0.093  &  &  	 0.094  	     &   0.096 &\\
$[$O/Fe$]$  &   $-$0.004   &  +0.018  &  +0.014 &  +0.058&  1   & 0.131  &  &  	 0.143  	     &   0.145 &\\
$[$Na/Fe$]$ &   $-$0.002   &$-$0.005  &$-$0.002 &  +0.042&  3   & 0.076  &  &  	 0.087  	     &   0.087 &\\
$[$Fe/H$]$I &     +0.005   &$-$0.002  &$-$0.004 &$-$0.042& 28   & 0.028  &  &  	 0.051  	     &   0.051 &NGC 3201\\
$[$Fe/H$]$II&   $-$0.002   &  +0.017  &  +0.010 &$-$0.017&  2   & 0.103  &  &  	 0.104  	     &   0.106 &\\
$[$O/Fe$]$  &   $-$0.004   &  +0.018  &  +0.019 &  +0.046&  1   & 0.146  &  &  	 0.153  	     &   0.155 &\\
$[$Na/Fe$]$ &   $-$0.002   &$-$0.004  &$-$0.000 &  +0.038&  2   & 0.103  &  &  	 0.110  	     &   0.110 &\\
$[$Fe/H$]$I &     +0.004   &$-$0.001  &$-$0.002 &$-$0.031& 12   & 0.037  &  &  	 0.048  	     &   0.048 &NGC 4590\\
$[$Fe/H$]$II&   $-$0.001   &  +0.015  &  +0.004 &$-$0.014&  1   & 0.129  &  &  	 0.130  	     &   0.131 &\\
$[$O/Fe$]$  &   $-$0.002   &  +0.016  &  +0.011 &  +0.039&  1   & 0.129  &  &  	 0.135  	     &   0.136 &\\
$[$Na/Fe$]$ &   $-$0.002   &$-$0.003  &  +0.002 &  +0.025&  3   & 0.074  &  &  	 0.078  	     &   0.078 &\\
$[$Fe/H$]$I &     +0.014   &$-$0.001  &$-$0.002 &$-$0.031& 33   & 0.024  &  &  	 0.042  	     &   0.042 &NGC 5904\\
$[$Fe/H$]$II&   $-$0.007   &  +0.018  &  +0.005 &$-$0.012&  2   & 0.098  &  &  	 0.099  	     &   0.101 &\\
$[$O/Fe$]$  &   $-$0.010   &  +0.018  &  +0.009 &  +0.031&  1   & 0.139  &  &  	 0.143  	     &   0.144 &\\
$[$Na/Fe$]$ &   $-$0.004   &$-$0.006  &$-$0.002 &  +0.021&  3   & 0.080  &  &  	 0.083  	     &   0.083 &\\
$[$Fe/H$]$I &     +0.004   &$-$0.001  &$-$0.001 &$-$0.037& 37   & 0.022  &  &  	 0.043  	     &   0.043 &NGC 6121\\
$[$Fe/H$]$II&   $-$0.003   &  +0.019  &  +0.006 &$-$0.003&  3   & 0.079  &  &  	 0.079  	     &   0.082 &\\
$[$O/Fe$]$  &   $-$0.003   &  +0.018  &  +0.009 &  +0.008&  1   & 0.136  &  &  	 0.136  	     &   0.138 &\\
$[$Na/Fe$]$ &     +0.000   &$-$0.007  &$-$0.003 &  +0.005&  3   & 0.079  &  &  	 0.079  	     &   0.080 &\\
$[$Fe/H$]$I &     +0.002   &$-$0.001  &  +0.001 &$-$0.055& 37   & 0.025  &  &  	 0.060  	     &   0.060 &NGC 6171\\
$[$Fe/H$]$II&   $-$0.001   &  +0.020  &  +0.012 &$-$0.027&  2   & 0.109  &  &  	 0.112  	     &   0.115 &\\
$[$O/Fe$]$  &   $-$0.001   &  +0.018  &  +0.014 &  +0.059&  1   & 0.154  &  &  	 0.165  	     &   0.166 &\\
$[$Na/Fe$]$ &     +0.000   &$-$0.008  &$-$0.001 &  +0.032&  3   & 0.089  &  &  	 0.095  	     &   0.095 &\\
$[$Fe/H$]$I &     +0.004   &$-$0.002  &$-$0.005 &$-$0.023& 25   & 0.026  &  &  	 0.035  	     &   0.035 &NGC 6254\\
$[$Fe/H$]$II&   $-$0.001   &  +0.016  &  +0.007 &$-$0.008&  2   & 0.091  &  &  	 0.091  	     &   0.093 &\\
$[$O/Fe$]$  &   $-$0.003   &  +0.018  &  +0.016 &  +0.026&  1   & 0.129  &  &  	 0.132  	     &   0.134 &\\
$[$Na/Fe$]$ &   $-$0.002   &$-$0.004  &  +0.000 &  +0.020&  3   & 0.074  &  &  	 0.077  	     &   0.077 &\\
$[$Fe/H$]$I &     +0.003   &  +0.007  &  +0.014 &$-$0.084& 19   & 0.037  &  &  	 0.092  	     &   0.093 &NGC 6388\\
$[$Fe/H$]$II&   $-$0.013   &  +0.025  &  +0.033 &$-$0.044&  2   & 0.114  &  &  	 0.123  	     &   0.130 &\\
$[$O/Fe$]$  &   $-$0.001   &  +0.011  &  +0.014 &  +0.086&  2   & 0.114  &  &  	 0.143  	     &   0.144 &\\
$[$Na/Fe$]$ &     +0.004   &$-$0.015  &$-$0.001 &  +0.027&  3   & 0.093  &  &  	 0.097  	     &   0.098 &\\
$[$Fe/H$]$I &     +0.004   &$-$0.001  &$-$0.001 &$-$0.031& 16   & 0.038  &  &  	 0.049  	     &   0.049 &NGC 6397\\
$[$Fe/H$]$II&   $-$0.001   &  +0.015  &  +0.003 &$-$0.010&  1   & 0.153  &  &  	 0.153  	     &   0.154 &\\
$[$O/Fe$]$  &   $-$0.003   &  +0.017  &  +0.010 &  +0.034&  1   & 0.153  &  &  	 0.157  	     &   0.158 &\\
$[$Na/Fe$]$ &   $-$0.002   &$-$0.002  &  +0.000 &  +0.031&  2   & 0.108  &  &  	 0.112  	     &   0.112 &\\
$[$Fe/H$]$I &     +0.006   &$-$0.002  &$-$0.005 &$-$0.030& 23   & 0.027  &  &  	 0.041  	     &   0.041 &NGC 6809\\
$[$Fe/H$]$II&   $-$0.002   &  +0.016  &  +0.006 &$-$0.008&  2   & 0.091  &  &  	 0.091  	     &   0.093 &\\
$[$O/Fe$]$  &   $-$0.004   &  +0.017  &  +0.015 &  +0.034&  1   & 0.128  &  &  	 0.132  	     &   0.134 &\\
$[$Na/Fe$]$ &   $-$0.003   &$-$0.006  &$-$0.001 &  +0.024&  2   & 0.091  &  &  	 0.094  	     &   0.094 &\\
$[$Fe/H$]$I &     +0.005   &  +0.001  &  +0.002 &$-$0.032& 37   & 0.023  &  &  	 0.040  	     &   0.040 &NGC 6838\\
$[$Fe/H$]$II&   $-$0.004   &  +0.020  &  +0.011 &$-$0.016&  3   & 0.080  &  &  	 0.082  	     &   0.085 &\\
$[$O/Fe$]$  &   $-$0.005   &  +0.017  &  +0.010 &  +0.033&  2   & 0.098  &  &  	 0.104  	     &   0.105 &\\
$[$Na/Fe$]$ &   $-$0.001   &$-$0.009  &$-$0.001 &  +0.013&  3   & 0.080  &  &  	 0.081  	     &   0.082 &\\
$[$Fe/H$]$I &     +0.005   &$-$0.002  &$-$0.001 &$-$0.026& 13   & 0.030  &  &  	 0.040  	     &   0.040 &NGC 7078\\
$[$Fe/H$]$II&   $-$0.001   &  +0.015  &  +0.004 &$-$0.007&  1   & 0.111  &  &  	 0.111  	     &   0.112 &\\
$[$O/Fe$]$  &   $-$0.003   &  +0.016  &  +0.011 &  +0.036&  1   & 0.111  &  &  	 0.117  	     &   0.118 &\\
$[$Na/Fe$]$ &   $-$0.002   &$-$0.003  &  +0.003 &  +0.023&  2   & 0.078  &  &  	 0.081  	     &   0.081 &\\
$[$Fe/H$]$I &     +0.005   &$-$0.001  &$-$0.001 &$-$0.033& 12   & 0.034  &  &  	 0.048  	     &   0.048 &NGC 7099\\
$[$Fe/H$]$II&   $-$0.001   &  +0.015  &  +0.003 &$-$0.008&  1   & 0.119  &  &  	 0.119  	     &   0.120 &\\
$[$O/Fe$]$  &   $-$0.002   &  +0.016  &  +0.007 &  +0.041&  1   & 0.119  &  &  	 0.126  	     &   0.127 &\\
$[$Na/Fe$]$ &   $-$0.002   &$-$0.003  &  +0.001 &  +0.025&  2   & 0.084  &  &  	 0.088  	     &   0.088 &\\
\\
\hline
\end{tabular}
\label{t:errabu}
\end{table*}

\end{appendix}

\end{document}